\documentclass{WileyMSP-template}

\begin{document}

\pagestyle{fancy}
\rhead{\includegraphics[width=2.5cm]{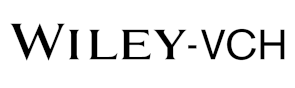}}

\title{3D printed waveguides for optogenetics applications: design optimization and optical characterization}

\maketitle

% Author: Please give full first and last names for authors and include * after the name of all corresponding authors

\author{Giorgio Scordo*}
\author{Kostas Kanellopulos}
\author{Surangrat Thongkorn}
\author{Samuel Tavares da Silva Maraschin}
\author{Kambiz Ghaseminasab}
\author{Evgeniy Shkondin}
\author{Deepshika Arasu}
\author{Stephan Sylvest Keller}
\author{Arto Rainer Heiskanen}
\author{Marta Perez Pereira}
\author{Jenny Emnéus*}

% Affiliations: Please provide adacemic titles (Prof. or Dr.) for all authors where applicable, and include an institutional email address for all corresponding authors
\begin{affiliations}
Dr. G. Scordo\\
Electronic Engineering Department, University of Rome Tor Vergata, Via del Politecnico 1, 00133 Rome, Italy\\
Email Address: giorgio.scordo@uniroma2.it

Dr. K. Kanellopulos\\
TU Wien, Institute of Sensor and Actuator Systems, Gusshausstrasse 27-29, 1040 Vienna, Austria

Dr. S. Thongkorn\\
Department of Biotechnology and Biomedicine, Technical University of Denmark, Kgs. Lyngby, 2800, Denmark

S. T. D. S. Maraschin\\
Department of Biotechnology and Biomedicine, Technical University of Denmark, Kgs. Lyngby, 2800, Denmark

K. Ghaseminasab\\
Department of Biotechnology and Biomedicine, Technical University of Denmark, Kgs. Lyngby, 2800, Denmark

Dr. E. Shkondin\\
Department of Biotechnology and Biomedicine, Technical University of Denmark, Kgs. Lyngby, 2800, Denmark

D. Arasu\\
Hospital Clínic de Barcelona - Fundació de Recerca Cliníc Barcelona - IDIBAPS - University of Barcelona, Barcelona, Catalonia, Spain

Prof. S.S. Keller\\
National Centre for Nano Fabrication and Characterization (DTU Nanolab), Technical University of Denmark, Ørsteds Plads Building 347, Kgs. Lyngby, 2800 Denmark

Dr. A. R. Heiskanen\\
Department of Biotechnology and Biomedicine, Technical University of Denmark, Kgs. Lyngby, 2800, Denmark

Prof. M. P. Pereira\\
The Institute for Molecular Biology of the Autonomous University of Madrid, Madrid, 20049, Spain

Prof. J. Emnéus\\
Department of Biotechnology and Biomedicine, Technical University of Denmark, Kgs. Lyngby, 2800, Denmark\\
Lund University, Department of Chemistry, Center for Analysis and Synthesis, Box 124, 22100, Lund, Sweden\\
Email Address: jenny.emneus@chem.lu.se

\end{affiliations}

% Keywords: Please provide a minimum of three and a maximum of seven keywords, separated by commas


% Abstract should be written in the present tense and impersonal style (i.e., avoid we), and be at most 200 words long
\begin{abstract}

Optogenetics has emerged as a powerful tool for disease modeling, enabling precise control of cellular activities through light stimulation and providing a valuable insights into disease mechanisms and therapeutic possibilities. Innovative materials and technologies such as micro-LEDs, optical fibers and micro/nano probes have been developed to allow precise spatial and temporal control of light delivery to target cells. Recent advances in 3D printing have further enhanced optogenetic applications by enabling the fabrication of implantable, customizable, and miniaturized light stimulation systems with high spatial resolution. In this study, we introduce a novel concept of a 3D printed light delivery system for brain organoid stimulation exploring the capabilities of projection microstereolithography (PµSL).
We characterized the optical properties of the high-resolution acrylate-based 3D print resin, i.e., refractive index and extinction coefficient, to evaluate if the light transmission efficiency might limit the performance of the optogenetic stimulation systems. Finite element method simulations were employed to optimize the  3D printed design.
An optogenetic setup was developed for optimal light delivery, and initial tests with optogenetically modified cells showed light-induced dopamine release with a stimulation efficiency of 2.8\%, confirming the 3D printed waveguide functionality and guiding future optimization. Our results demonstrate that this light stimulation tool offers strong potential for advancing customizable optogenetic applications.

\end{abstract}

% Text: Please use section headings and subheadings as specified below. For communications, all section headings apart from Experimental Section should be removed
% Please make the first reference to a display item bold: \textbf{Figure 1}
% Do not abbreviate Figure, Equation, etc.; display items are always singular, i.e., Figure 1 and 2.
% Equations are always singular, i.e., Equation 1 and 2, and should be inserted using the {equation} environment, not as graphics
% Please do not use footnotes in the text, additional information can be added to the Reference list.

\section{Introduction}
\begin{figure}
  \includegraphics[width=\linewidth]{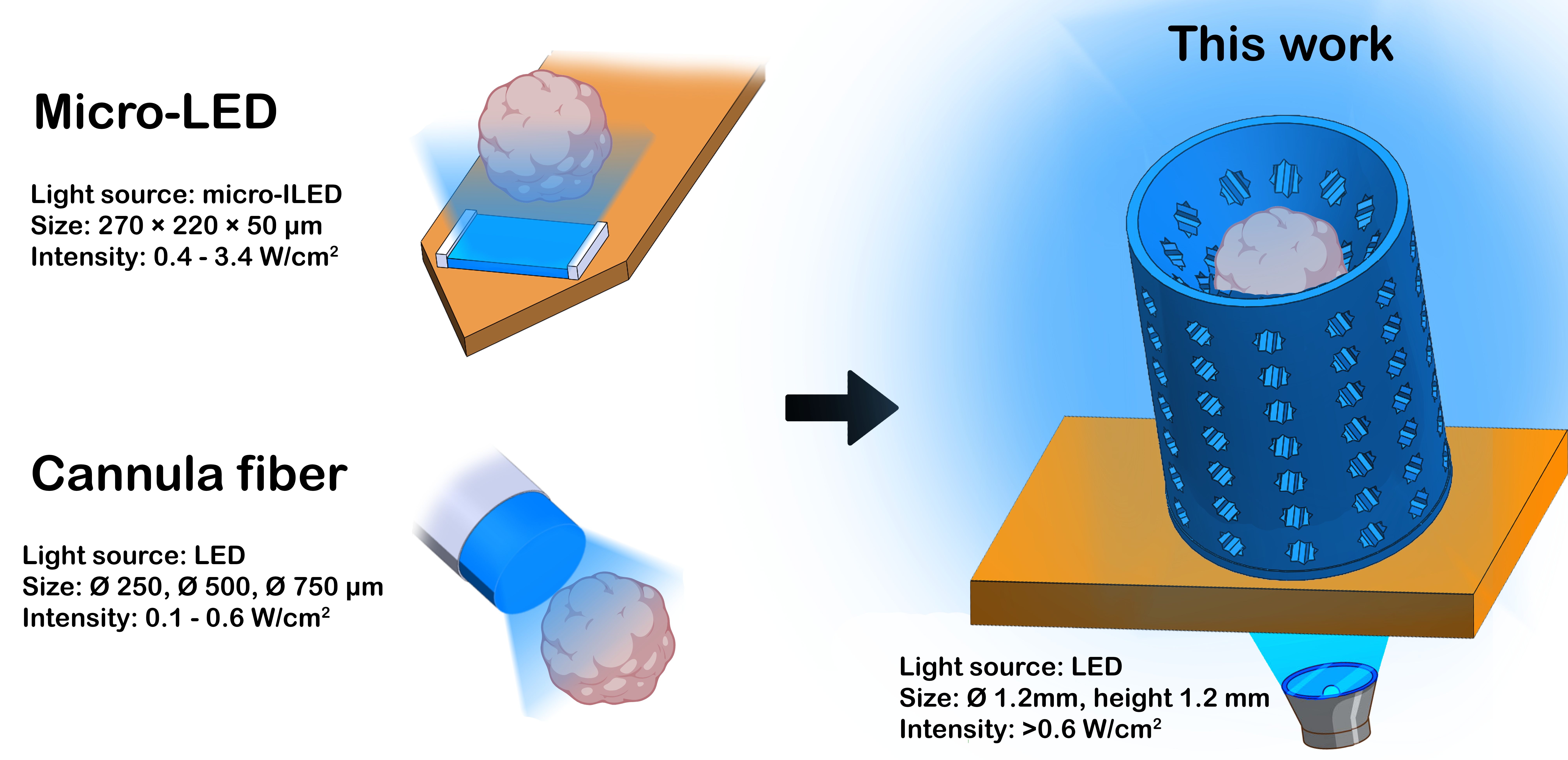}
  \caption{Schematic representation of organoid photostimulation systems. (Left) 2D photoexcitation approaches: (top) full implanted Micro-LED; (bottom) partially implanted cannula coupled to an optical fiber. (Right) Innovative 3D optogenetic stimulation system proposed in this work for 3D optogenetic cells and organoids stimulation.}
  \label{fig:Intro}
\end{figure}

Optogenetics is a technique that integrates optics and genetics to control the activity of specific cells in living tissue, typically neurons \cite{Chen2022} or cardiac cells \cite{leemann2023cardiac}, with high spatial and temporal precision. This method relies on light-sensitive proteins, such as channelrhodopsins \cite{tsukamoto2021optogenetic} and halorhodopsins \cite{parrish2023activation}, which can be genetically introduced into cells to enable temporally controlled photostimulation. Optogenetics has revolutionized neuroscience by allowing researchers to investigate neural circuits, understand brain function, and develop potential therapeutic applications for neurological 
disorders  \cite{zhang2024application}.
For an optogenetic system to function effectively, several key parameters must be carefully considered, including  wavelength, intensity, divergence, and penetration depth of light \cite{abaya20123d, yona2016realistic}. These factors collectively influence the efficiency of light propagation within the target tissues, the precision of stimulation, and the overall dimensions of the implanted system. Building on these considerations, a variety of device architectures have been developed to deliver optical stimulation in vivo, most commonly based on optical fibers coupled to external light sources or micro-LED-based implants, which have proven reliable and user-friendly (\textbf{Figure \ref{fig:Intro}}) \cite{sparta2012construction, jeong2015wireless, silva2025wireless, ajieren2024design} . Among these, optical fiber–based systems remain the most established approach for delivering light to deep brain regions, offering robust and controllable illumination; however, their rigid geometry and intrinsic light scattering within biological tissue limit spatial precision and uniform light distribution, particularly in volumetric or heterogeneous 3D models \cite{yona2016realistic, zhang2022emerging}. In contrast, micro-LED-based systems offer improved spatial selectivity, mechanical flexibility, and reduced invasiveness, representing a promising alternative for chronic or localized optogenetic stimulation \cite{zhang2022emerging}. Advanced designs, including tapered and multifunctional fibers integrating micro-optical structures \cite{pisanello2018tailoring} wireless implantable micro-LED probes \cite{mondello2021micro} and and thin-film microscale LED platforms \cite{zhang2022emerging} have partially mitigated these drawbacks by improving light homogeneity, spatial control, and reducing shadowing effects. Despite these advancements, current systems remain limited when applied to three-dimensional (3D) tissue environments, where optogenetic models such as spheroids or organoids are implanted, as restricted light penetration and proximity constraints hinder homogeneous and efficient photostimulation. Addressing these challenges requires not only advances in device architecture but also in the optical materials used for light delivery. Therefore, materials selected for constructing waveguides capable of stimulating 3D tissue models must exhibit a well-defined refractive index, low optical absorption in the visible range, high transparency, and biocompatibility to achieve efficient and reproducible photostimulation \cite{farnham2014optogenetics}.

Several materials have been investigated for their suitability in optogenetic applications, each with distinct advantages and limitations. Polydimethylsiloxane (PDMS) is widely used due to its flexibility and biocompatibility; however, its high light absorption and low refractive index limit its effectiveness for optical applications \cite{rudmann2024fabrication}. Parylene-C offers excellent biocompatibility and moisture resistance, but it poses challenges in structuring complex 3D geometries, restricting its adaptability for intricate optogenetic systems \cite{reddy2023vivo}. Poly(methyl methacrylate) (PMMA) provides high optical clarity, yet its brittleness and mechanical instability hinder its long-term reliability \cite{frank2022optofluidic}.

One of the most used materials is SU-8 \cite{chen2024implantation}, which is valued for its high refractive index for light confinement ($\sim 1.63$ at 470 nm, $\sim 1.60$ at 590 nm) \cite{major2021spectroscopic}, and its low absorption coefficient of $\sim 0.002$ at 470 nm \cite{ashraf2019determination}. These optical properties make it an excellent candidate for waveguide applications. However, SU-8 is an epoxy-based negative photoresist that must be processed in a cleanroom via photolithography and is, due to its multi-stage curing, incompatible with methods intended to form high-aspect-ratio 3D architectures.

Recent advances in 3D printing technologies have introduced novel materials that could overcome the aforementioned limitations \cite{zhu2022recent}. The ability to fabricate complex 3D optogenetic system using micro-scale additive manufacturing allows for precise control of light propagation and spatial distribution. Several 3D-printable materials have been explored for optogenetic stimulation: Methacrylate-based resins offer high optical transparency and mechanical stability for implantable optogenetic devices \cite{monney2021photolithographic}. Poly(ethylene glycol) diacrylate (PEGDA)-based hydrogels are biocompatible with tunable optical properties and allow light-guiding in biological environments \cite{feng2020printed}. Silicon-based hybrid materials combine the advantages of polymers and inorganics to enhance optical performance in neural interfaces \cite{mohanty2018reconfigurable}. 
Although several photopolymer materials have been reported for optogenetic applications, acrylate-based resins remain largely unexplored, and their optical behavior and waveguide design principles are poorly understood. In this study, we investigated the fabrication of a three-dimensional optical waveguide using projection microstereolithography (PµSL) to enable controlled light delivery to optogenetically modified neuronal cells. The Boston Micro Fabrication (BMF) high-resolution acrylate resin was selected and systematically characterized in terms of refractive index, extinction coefficient, and light transmission efficiency to evaluate its suitability for optical stimulation. Based on these data, a 3D optical system was designed and optimized through Finite Element Method (FEM) simulations, and an optogenetic setup incorporating a custom-built Optical Motherboard was developed to assess light delivery performance (Figure \ref{fig:Intro}, right). To validate the functionality of the 3D-printed system, a proof-of-concept experiment was conducted in which light transmitted through the BMF waveguide was used to stimulate optogenetically modified dopaminergic cells differentiated on pyrolytic carbon fibers  \cite{vasudevan2019leaky}, positioned within the 3D printed cavity. Light-induced dopamine release was quantitatively detected on the carbon fibers acting as working electrodes, confirming the effectiveness of the 3D printed  waveguide for precise light transmission and functional photostimulation.

\section{Methods}
\subsection{Optical properties of BMF resin}
To systematically assess the suitability of biocompatible HTL Yellow 20 µm resin from Boston Micro Fabrication (BMF), USA, for optogenetic applications, we conducted a comprehensive investigation of its optical properties, quantifying its refractive index ($n$) and extinction coefficient ($\kappa$), which serve as key inputs for realistic FEM simulations. Additionally, measurements of transmitted light intensity and divergence were performed to validate the performance of the light source during the 3D-printed optical waveguide testing.

\subsubsection{Refractive index and extinction coefficient}
Ellipsometry was used to characterize the refractive index ($n$) of the BMF resin in the wavelength range from 300 to 1000 nm. The measurements were carried out using disks 3D printed (\textbf{Figure \ref{fig:Opt_char}a}) with PµSL 3D printer (microArch® S140, BMF, USA). The optical dispersion $n(\lambda)$ of the resin was measured using an Angle Spectroscopic Ellipsometer M-2000XI (J.A. Woollam, US). In ellipsometry, key optical parameters are determined via analysis of the changes in the light polarization upon reflection. The measurements were performed at six different angles of incidence (45°-70°, with steps of 5°), to enhance the precision of extracted optical parameters \cite{fujiwara2007spectroscopic}.

The extinction coefficient ($\kappa$) was determined with spectrophotometry due to its higher absorption sensitivity compared to ellipsometry ($\kappa\sim 10^{-3}$) \cite{tompkins2005theory, corato2024absorption}. For this purpose, a 3D printed BMF samples was designed in the form of a cuvette with dimensions of 1 cm × 1 cm × 5 cm (see supplementary information S2A).
Spectrophotometry is a widely used optical technique for determining the refractive index and extinction coefficient of materials by analyzing their absorbance and transmittance across different wavelengths \cite{paul2019synthesis, bdewi2016synthesis}. In this study, a UV-Vis-NIR Evolution Spectrophotometer 201/220 (Thermo Fisher Scientific, US) was employed to measure the absorbance $A=-\mathrm{log}_{10}(T)$ of the resin sample of thickness $h=1$ cm, over a wavelength range of 400-800 nm, where $T=I_t/I_0$ is the experimental optical transmittance, with $I_0$ and $I_t$ being the input and transmitted intensity, respectively. From the absorbance, it is possible to determine the extinction coefficient according to \cite{macleod2010thin}
\begin{equation}\label{eq:kappa}
    \kappa(\lambda) = \frac{\lambda \alpha(\lambda)}{4 \pi},
\end{equation}
where $\lambda$ is the probing wavelength, and $\alpha(\lambda)=A(\lambda)/h$ is the absorption coefficient of the material, given in units [1/m].

\subsubsection{Intensity and divergence of the transmitted light}
\begin{figure}
  \includegraphics[width=\linewidth]{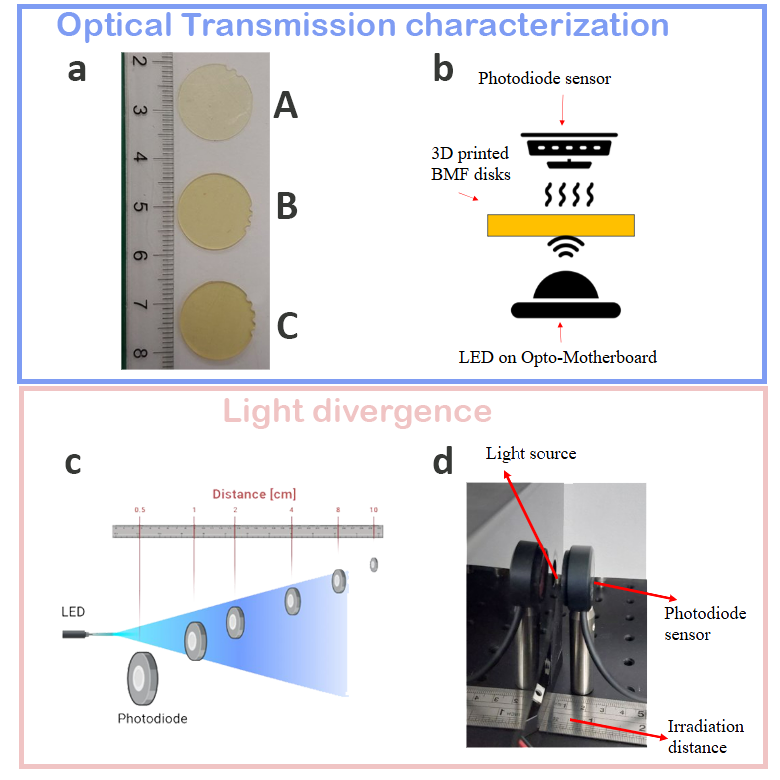}
  \caption{\textbf{a-b} Optical transmission characterization setup. \textbf{a} 3D printed disks with different thickness: $h_\mathrm{A}=0.5$ mm, $h_\mathrm{B}=1.0$ mm, $h_\mathrm{C}=1.2$ mm. \textbf{b} A schematic of the experimental setup. \textbf{d-e} Light divergence characterization setup. \textbf{a} A schematic view of the distribution and the expansion of the light beam. \textbf{b} A lateral view of the experimental setup.}
  \label{fig:Opt_char}
\end{figure}

The transmitted intensity ($I_t$) quantifies the amount of optical power that passes through a material after partial absorption and scattering. It is influenced by key factors such as the absorption coefficient $\alpha(\lambda)$, material thickness $h$, and the intensity of the incident light $I_0$. 
To determine $I_t$, three disks (A, B, and C) with a diameter of 15 mm and thicknesses of 0.5, 1, and 1.2 mm, respectively, were fabricated using BMF resin (Figure \ref{fig:Opt_char}a). The disks were subsequently placed in a plastic holder between the LED light source (LED470 with blue light and LED590 with orange light, Thorlabs, USA) and the photodiode sensor (S121C Thorlabs, USA) connected to a power meter (PM121D Thorlabs, USA) as illustrated in Figure \ref{fig:Opt_char}b. The LED was kept continuously emitting for 5 min to assess potential fluctuations in light output, while the transmitted intensity through the material was recorded over this period using the photodiode.

The light divergence governs the spatial distribution and expansion of the light beam as it propagates through a medium. For the present application, high divergence of the light source would be detrimental, as it will reduce the intensity available for photostimulation. To quantify its contribution to the overall performance in light delivery, the divergence angle of the LEDs used in this work was characterized in terms of the half-power angle $\theta_{1/2}$, i.e., the zenith angle with respect to the optical axis that defines a cone containing 50$\%$ of the power emitted by the LED source \cite{kashiwao2025modeling}. It was measured with the photodiode positioned at distances of 0.5, 1, 2, 4, 8, and 10 cm from the LED source with no BMF resin structure in between, as shown in Figure \ref{fig:Opt_char}c. The measurements were carried out in the dark (see Figure \ref{fig:Opt_char}d) with a light exposure of 5 min for each position.

\subsection{Optogenetic setup}\label{Opt_setup}
\begin{figure}
  \includegraphics[width=1.\linewidth]{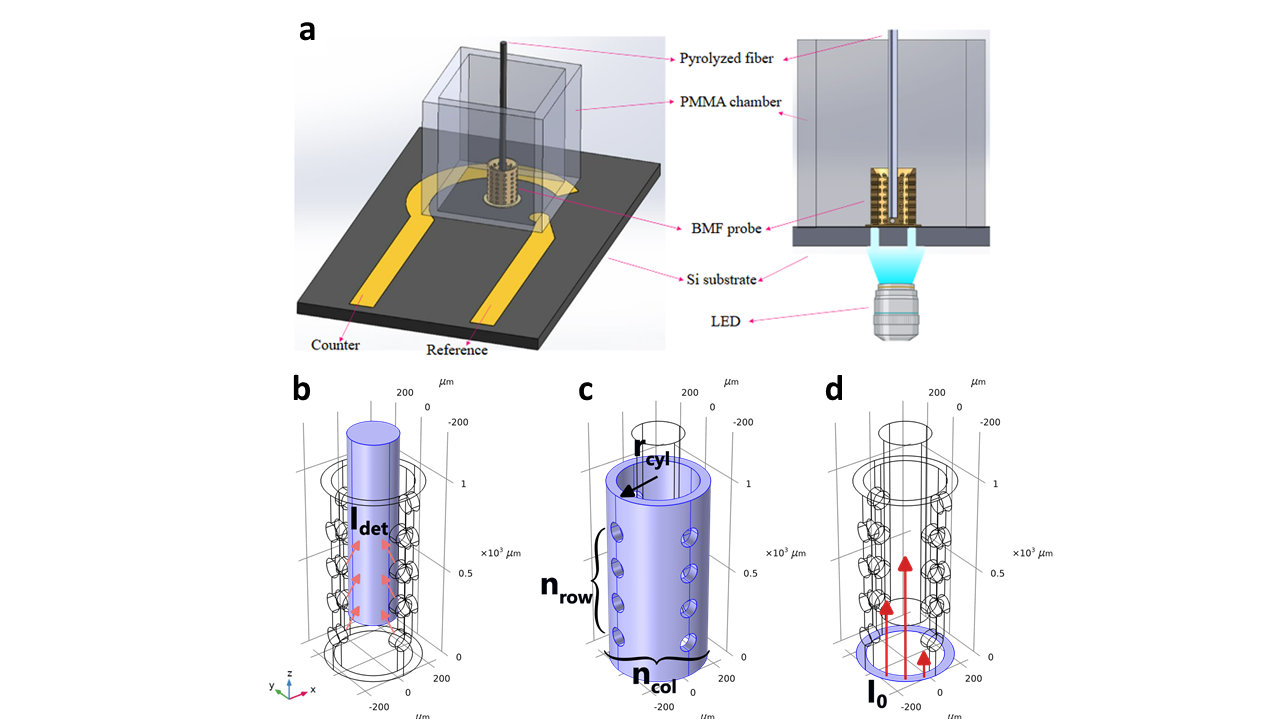}
  \caption{Experimental setup and FEM modeling. \textbf{a} Schematic sketch of the optogenetic system testing . \textbf{b} Definition of the power detector (light red arrows indicate the scattered rays absorbed and measured by the detector). \textbf{c} Design of the device, with a focus on the main geometrical parameters used for optimization. \textbf{d} Definition of the light source at the bottom surface of input intensity $I_0$ (the red arrows indicate the directionality of the emission).}
  \label{fig:BMF_setup_FEM}
\end{figure}

To assess the suitability of the optogenetic stimulation systems proposed in this work for optogenetic applications, we designed and implemented a specialized experimental setup, as depicted in \textbf{Figure \ref{fig:BMF_setup_FEM}a}. This setup integrates multiple fabrication techniques, including cleanroom processing, laser micromachining, and high-precision 3D printing, to create a functional optogenetic waveguide, featuring a hollow cylindrical geometry with several openings patterned in its walls (Figure \ref{fig:Intro}, right; Figure \ref{fig:BMF_setup_FEM}a).

First, a 2D electrochemical sensor was fabricated on a silicon wafer using a cleanroom fabrication process (for details see supporting information SI, Section~S4), resulting in gold reference and counter electrodes (see Figure \ref{fig:BMF_setup_FEM}a). Subsequently, the silicon chips electrodes were engraved obtaining precise aperture on either side of the electrode, allowing light to traverse the silicon substrate and couple efficiently into the 3D-printed waveguide structure (for more details see SI, Section~S3C).
The cylindrical structure, optimized through FEM simulation (Section \ref{FEM_settings}), was designed using the CAD software SolidWorks (SolidWorks, USA) and exported as an STL file. The file was then sliced by the software ChituBox (ChituBox, China) using 5 µm layer thickness and printed using the BMF PµSL 3D printer adjusted to 1 s exposure time at 45$\%$ intensity (for more details see SI, Section~S3C). The 3D printed waveguide structure was precisely aligned over the working area of the electrode, ensuring that the surrounding wall of the cylinder was positioned above the cavity. This design lets light pass through the material, scatter off the patterned structures, and reach the optogenetic cell culture on the pyrolyzed fiber (Figure \ref{fig:BMF_setup_FEM}a, right).

\subsection{FEM modeling}\label{FEM_settings}
Finite element simulations (COMSOL® Multiphysics, v5.5) were carried out to determine the impact of the geometrical parameters on the optical intensity of the transmitted light from the cylindrical 3D printed waveguide to its central axis, available for photostimulation ($I_{det}$, light red arrows in Figure \ref{fig:BMF_setup_FEM}b). The Geometrical Optics Physics was exploited with a focus on the Ray Tracing Study, as the geometrical features of the device are much larger than the optical wavelength ($\gg \lambda$). Different 3D waveguides  were explored in terms of the geometry of the openings (Figure \ref{fig:BMF_setup_FEM}c), with the number of rows of holes, $n_{row}$, number of columns, $n_{col}$, and cylinder's external radius, $r_{cyl}$, being the optimized parameters. For all the different combinations, the height of the cylinder ($h_{cyl}=1.2$ mm---for more info about this choice, see \textbf{Figure \ref{fig:LED_div}d} and relative discussion), radius of the hole ($r_{hole}=50$ µm), and thickness of the wall ($t_{cyl}=50$ µm) were kept fixed. For each simulation, a light source with uniform intensity distribution was defined at the bottom of the probe, from which the rays are released with a conical directional vector pointing along the z-axis with a conical angle of 7\textdegree (red arrows in Figure \ref{fig:BMF_setup_FEM}d). The total input power was set to $P_0 = 0.1$ W to match the experimental values used in this work. The rays travel within the cylindrical probe and are partially transmitted toward its central axis (light red arrows in Figure \ref{fig:BMF_setup_FEM}b). The transmitted intensity available for photostimulation $I_{det}$ is then evaluated with a cylindrical ray detector positioned along the central axis (representing the pyrolyzed fiber of Figure \ref{fig:BMF_setup_FEM}a, right, with a radius $r_{det}=125$ µm and height $h_{det}= 1.2$ mm, at a distance of $h_{cyl}/4$ from the bottom), and evaluated as the ratio $P_{det}/A_{det}$ between the optical power impinging on the detector $P_{det}$ and its useful area $A_{det}=2\pi r_{det} h_{det} + \pi r_{det}^2$ (for more details about the distribution of the optical power on the detector, see SI S1). An optimization of the mesh was carried out upon a convergence analysis with a selected maximum element size of $t_{cyl}/2$ (with $t_{cyl}$ being the smallest geometry feature). For the optical properties of the 3D printed waveguide, the experimental results described in Section~\ref{Opt_resin} have been used as input parameters.

\subsection{Culturing of optogenetically modified cells}
10 cm long segments of optical fiber with fused silica core (\#57-086, Edmund Optics Ltd., UK) were pyrolyzed (Figure \ref{fig:BMF_setup_FEM}a, right) and subsequently treated with oxygen plasma as previously described to enhance the wettability of the pyrolytic carbon \cite{vasudevan2019leaky}. Following plasma treatment, the fibers were coated with poly-L-lysine (PLL-P6282, Merck Sigma-Aldrich, US) at a concentration of 50 µg mL$^{-1}$ in PBS for 2 h in an incubator. Channelrhodopsin-2 (ChR2) modified human ventral mesencephalic neural stem cells (hVM1-Bcl-XL-ChR2-mCherry) \cite{Vasudevan2023, vasudevan2019leaky} were then seeded at a pipetting 50 µL of a cell suspension (cell density of $2\cdot10^6$ cells mL$^{-1}$) per fiber (only a 5 mm section of each fiber exposed during cell seeding) and cultured for 24 h in growth medium containing DMEM/F12 with Gluta-MAX (Thermo Fisher Scientific Inc., USA), supplemented with 6 g L$^{-1}$ glucose and non-essential amino acid solution (100x dilution) from Merck, USA, 5 mM HEPES, 0.5\% AlbuMAX, N-2 supplement (100x dilution), and penicillin/streptomycin (100x dilution) all acquired from Thermo Fisher Scientific Inc., USA, as well as 20 ng L$^{-1}$ each of Recombinant Human Epidermal Growth Factor (EGF) and Recombinant Human Fibroblast Growth Factor (FGF) from Bio-Techne Ltd., USA. After 24 hours, cell differentiation was initiated by substituting the EGF and FGF with 1 mM dibutyryladenosine 3’, 5’-cylic monophosphate sodium salt (Merck, USA) and 2 ng mL$^{-1}$ recombinant human glial cell line-derived neurotrophic factor (GDNF) (Bio-Techne Ltd., USA) and continued for 10 days. During culturing and differentiation, the cell loaded carbon fibers were maintained in a humidified incubator at 37 °C under 95\% air/5\% CO$_2$.

\subsection{Neurotransmitter measurement}
The ability of the differentiated hVM1-Bcl-X\textsubscript{L}-ChR2-mCherry cells to release dopamine upon light-induced and chemical stimulation was evaluated by using chronoamperometry. The electrochemical measurement setup comprised a 3-electrode configuration, with a pyrolytic-carbon-coated optical fiber as the working electrode (WE), a gold counter electrode (CE), and a reference electrode (RE), all connected to a CH Instruments 1010 potentiostat (USA). During the recording of the baseline and dopamine release, the carbon fiber WE was poised at 0.5 V vs. the RE. During the measurements, the optical fiber was inserted into the lumen of the 3D printed waveguide under the guidance of a micromanipulator. The silicon chip, on which the optogenetic waveguide had been 3D printed, was placed in a micromilled vial fabricated of PMMA. For photostimulation of cells, light pulses were triggered using an Optical Motherboard (OpM) equipped with a blue light emitting LED (470 nm) (for details, see SI, Section S3A). As a control, chemical stimulation of the cells was conducted by pipetting high-K+ differentiation medium (additional 450 mM of KCl) into the chamber, where the 3D printed waveguide and cell loaded optical fiber were placed, to raise the K+ concentration to $\sim 150$ mM.

\section{Results and discussion}
The optogenetic 3D  stimulation systems designed and characterized in this study features a hollow cylindrical geometry, with openings patterned on its lateral surface (Figure \ref{fig:Intro}, right; Figure \ref{fig:BMF_setup_FEM}a). This geometry allows for central placement of the target cell, ensuring symmetric and uniform light delivery from the surrounding openings. The perforations in the 3D printed waveguide wall not only permit an efficient light delivery to the cell during photostimulation, but also provide fluidic exchange routes, enabling simultaneous light stimulation and sampling. Such integration is particularly advantageous for minimally invasive neuromodulation applications, where both controlled light delivery and fluid handling are essential. 

\subsection{Optical properties of BMF resin}\label{Opt_resin}
\subsubsection{Refractive index and extinction coefficient}
The optical properties of materials used in optogenetic stimulation systems are fundamental for light propagation and interaction within optogenetic culture targets. In particular, the refractive index ($n$) and the extinction coefficient ($\kappa$) play pivotal roles. The refractive index dictates the speed, direction, and dispersion of light as it traverses a medium, directly influencing the optical confinement, waveguiding efficiency, and mode propagation in optoelectronic and photonic applications. In contrast, the extinction coefficient accounts for the optical attenuation due to the absorption and scattering of power by the material \cite{reider2016photonics, macleod2010thin}. These parameters are critical in determining the efficiency and precision of optogenetic stimulation.

To the best of our knowledge, the $n$ and $\kappa$ of the BMF 3D printed resin at the key optogenetic wavelengths of 470 nm and 590 nm have not previously been reported in the literature.

\begin{figure}
  \includegraphics[width=\linewidth]{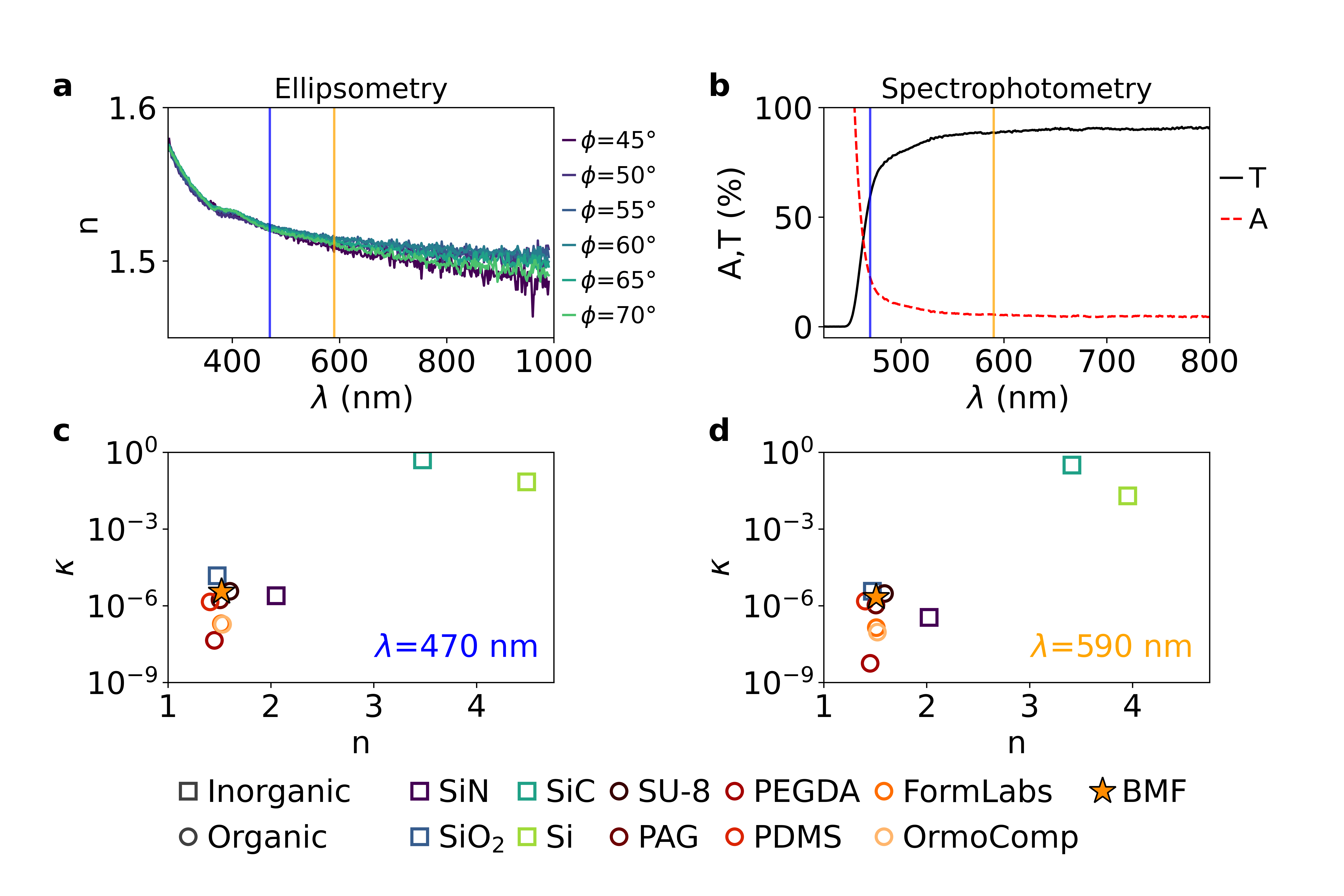}
  \caption{Optical characterization of BMF 3D print resin and comparison to other materials. \textbf{a} Ellipsometric determination of $n$ as a function of the excitation wavelength $\lambda$ for different incident angles $\phi$. \textbf{b} Spectrophotometric measurements of light transmission $T$ as a function of $\lambda$ (black solid curve) together with the corresponding calculated absorbance $A$ (red dashed curve). \textbf{c}-\textbf{d} Comparison of the BMF resin ($n$, $\kappa$) with the other materials exploited in optogenetics at 470 nm (c) and 590 nm (d) probing wavelength.}
  \label{fig:BMF_Opt_data}
\end{figure}

Ellipsometry measurements were carried out using PµSL 3D printed disks (Figure \ref{fig:Opt_char}a) to determine the refractive index $n$ of the BMF resin in the 300-1000 nm wavelength range. The experimental results in \textbf{Figure \ref{fig:BMF_Opt_data}a} show spectra of $n$ measured at different incident angles $\phi$ between the normal to the sample surface and the impinging light beam. As can be observed, the results are consistent, presenting minimal variation across the various angles of incidence, further highlighting the stability of the optical properties of the resin. Based on these measurements, the determined refractive indices at the two main wavelengths of interest, 470 nm (blue light) and 590 nm (orange light), were $n=1.52$ and $n=1.51$, respectively, similar to what reported for other 3D print resins \cite{reynoso2021refractive, gissibl2017refractive}.

Spectrophotometric measurements in the wavelength range 400-800 nm were carried out using cuvettes filled with cured  BMF resin to determine $\kappa$ at 470 nm and 590 nm. Figure \ref{fig:BMF_Opt_data}b shows the measured transmittance $T$ (black solid curve) and the calculated absorbance $A$ (red dashed curve)
\begin{equation}\label{eq:absorbance}
    A(\lambda)=-\mathrm{log}[T(\lambda)],
\end{equation}
as a function of $\lambda$. The BMF resin shows an exponential reduction in absorbance $A$ with increasing $\lambda$, which is consistent with its formulation for 3D printing at around 405 nm. This resin is design to absorb strongly at this wavelength due to the presence of photoinitiators, which confine polymerization to thin layers during the printing process.

Based on these results and calculations using Equation \ref{eq:kappa}, the determined extinction coefficients were $\kappa(470\ \mathrm{nm})=3.55\times 10^{-6}$ and $\kappa(590\ \mathrm{nm})=2.18\times 10^{-6}$.

The experimental results discussed above were compared with the complex refractive index ($n+i\kappa$) of other materials exploited in optogenetics at wavelengths of 470 nm and 590 nm, as shown in Figure \ref{fig:BMF_Opt_data}c and d. The plots present the imaginary part ($\kappa$) on a logarithmic scale on the y-axis, and the real part $n$ on the x-axis. At both wavelengths, BMF resin (represented by a star marker) exhibits a relatively high refractive index among the organic materials, closely approaching values typical of inorganic materials such as SiN and SiO$_2$. Importantly, BMF resin maintains an extremely low extinction coefficient ($\kappa\sim 10^{-6}$), indicating minimal optical absorption, which is a desirable characteristic for transparent optoelectronics devices. In comparison, inorganic materials like Si and SiC show significantly higher $\kappa$ values, particularly at 590 nm, where their optical losses reach up to three orders of magnitude greater than those of BMF resin. Organic materials, such as PEGDA, PDMS, and Polyalkylene (PAG), exhibiting low extinction coefficients, have considerably lower refractive indices (around $n\sim 1.4$ and 1.5), which may limit their utility in applications requiring high contrast in $n$. Notably, SU-8, another commonly used organic material for optical waveguide fabrication, shows similar optical transparency ($\kappa$ value) to BMF but with a slightly lower $n$. These findings position BMF resin as a promising candidate material for optogenetics applications.

\subsubsection{Transmitted light intensity and divergence}
In addition to the intrinsic optical properties of the material, the characteristics of the light source significantly influence the outcomes of optogenetic experiments. Light divergence affects beam expansion and focusing capabilities, directly impacting spatial resolution and targeted stimulation. This parameter is essential for the design of high-performance optogenetic stimulation systems, as it influences light penetration, scattering, and intensity distribution toward the target. Moreover, the transmitted intensity is determined by both the optical and geometrical properties of the material, i.e., absorption coefficient and thickness, and the light source parameters, including incident intensity and wavelength. The transmitted intensity ultimately defines the energy available for optogenetic protein activation and subsequent cellular responses.
These considerations are particularly crucial in optogenetics, where efficient light delivery is necessary for precise neural stimulation, especially in three-dimensional cell cultures such as organoids and spheroids.

To achieve efficient activation of optically sensitive proteins as those used for optogenetics \cite{Deisseroth2011}, it is crucial to optimize light delivery by minimizing the divergence while maximizing the transmitted intensity. We characterized the divergence properties of the light emitted from the two distinct LED sources (470 nm and 590 nm), without BMF in between, as well as the corresponding light transmission T through the BMF resin, analyzing how it is influenced by both the emission characteristics of the LEDs and the optical properties of the BMF resin.
\begin{figure}
  \includegraphics[width=\linewidth]{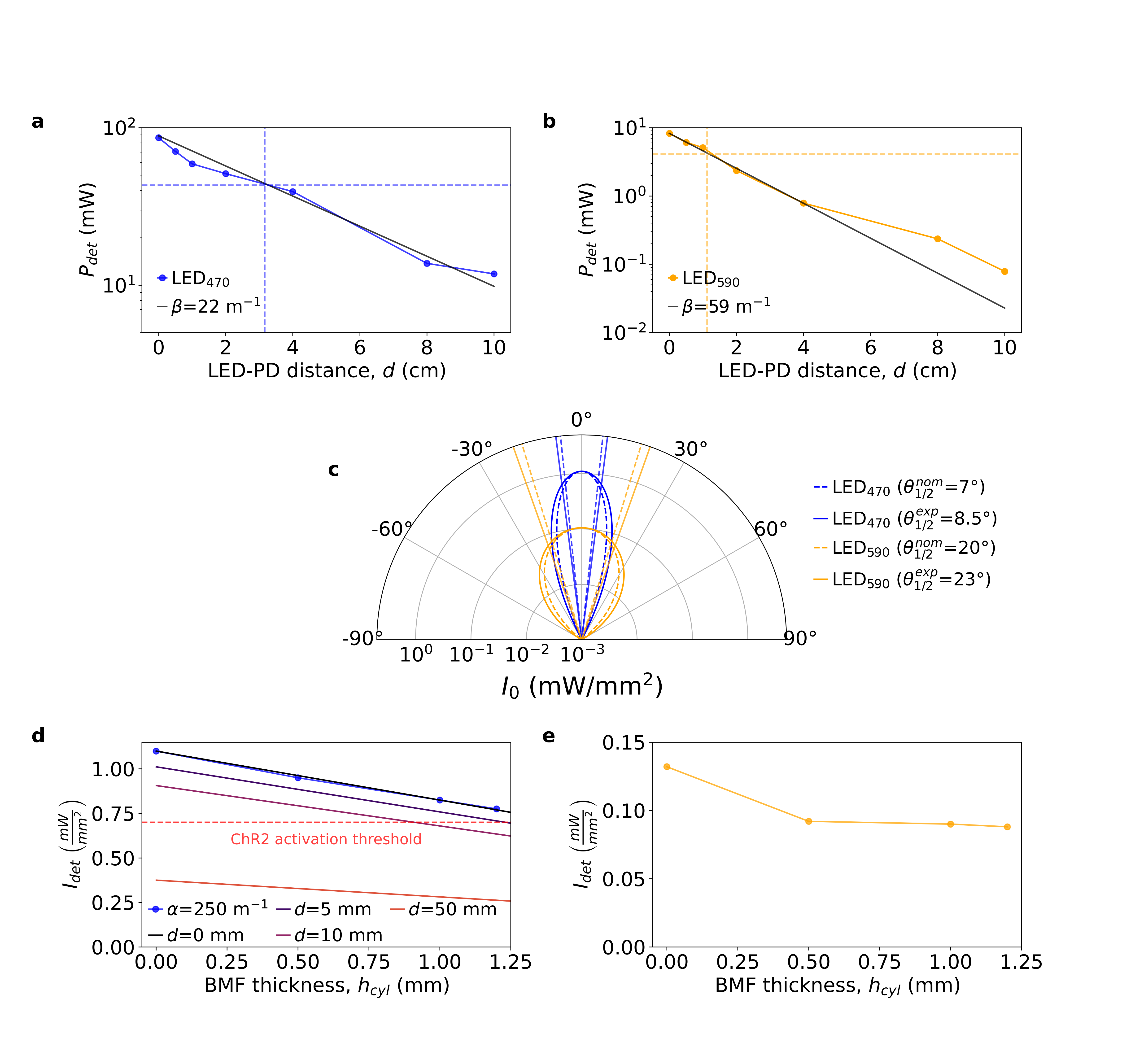}
  \caption{Characterization of light transmission and divergence. \textbf{a-b} Detected power $P_{det}$ emitted by the (\textbf{a}) LED$_{470}$ (blue curve) and (\textbf{b}) LED$_{590}$ (orange curve) as a function of the LED-photodetector distance ($d$), with no resin in between. The performed curve fitting (black solid curve) according to Equation \ref{eq:P_det} gives a divergence coefficient ($\beta$) of (\textbf{a}) $22$ m$^{-1}$ and (\textbf{b}) $59$ m$^{-1}$. The dashed lines indicate $P_{det}=P_0/2$ (horizontal) and the corresponding $d$ (vertical). \textbf{c} Comparison between the experimental (solid curves) and nominal (dashed curves) intensity profiles for the LED$_{470}$ (blue) and LED$_{590}$ (orange). Both profiles are Gaussian \cite{kashiwao2025modeling}. \textbf{d-e} Intensity attenuation for (\textbf{d}) LED$_{470}$ and (\textbf{e}) LED$_{590}$ as a function of the disk thickness $h_{cyl}$. In \textbf{d}, an absorption coefficient $\alpha=250$ m$^{-1}$ has been extracted based on linear regression. The horizontal dashed line indicates the threshold intensity for ChR2 photostimulation. Solid lines indicates the intensity reduction within the BMF disk, for different LED-to-disk distance $d$.}
  \label{fig:LED_div}
\end{figure}

As seen in Figure \ref{fig:LED_div}a and b, both LEDs exhibit an exponential decay of their emitted optical power with increasing LED-photodiode sensor distance $d$ according to
\begin{equation}\label{eq:P_det}
    P_{det}(d)=P_0 e^{-\beta d},
\end{equation}
from which a divergence coefficient $\beta$ of 22 m$^{-1}$ and 59 m$^{-1}$ has been extracted for the LED$_{470}$ and LED$_{590}$, respectively.

For LED$_{470}$, approximately 80$\%$ of its total emitted power $P_0$ can be detected within the first 5 mm (Figure \ref{fig:LED_div}a). For LED$_{590}$, this value reduces to 70$\%$ (Figure \ref{fig:LED_div}b). The small divergence at short distances is a sign of a strong directivity, i.e., reduced angular dispersion, of the LEDs. From the measurements in Figure \ref{fig:LED_div}a and b, the half-power angle $\theta_{1/2}$ of each LED was extracted according to
\begin{equation}\label{eq:div_angle}
    \theta_{1/2} = \arctan\left(\frac{r_{det}}{d_{1/2}}\right),
\end{equation}
where $r_{det}$ and $d_{1/2}$ are the radius of the detector's active area (here 4.75 mm), and the LED-photodiode distance at which the detected power is $P_{det}(d_{1/2})=P_0/2$. For the LED$_{470}$, a $\theta_{1/2}=8.54$\textdegree was found, closely aligning with the reported value of 7\textdegree \cite{led470}. For the LED$_{590}$, a $\theta_{1/2}=23$\textdegree was found, again closely aligning to the nominal value of 20\textdegree \cite{led590}. The corresponding intensity distribution are shown in Figure \ref{fig:LED_div}c (solid curves), indicating an excellent agreement with the nominal profiles (dashed curves). These results highlight the fact that the LEDs present an intensity distribution more similar to the Gaussian rather than Lambertian profile \cite{kashiwao2025modeling}.

In Figure \ref{fig:LED_div}d, the optical intensity of the LED$_{470}$ measured as a function of the thickness $h_{cyl}$ of the BMF resin disk samples, is displayed for an input intensity of $I_0=1.1$ mW mm$^-2$, with the LED and PD directly in contact with the disks ($d=0$ mm), exhibiting a monotonically decreasing trend with increasing disk thickness. Upon linear regression, an absorption coefficient $\alpha(470 \mathrm{nm})=250$ m$^{-1}$ was extracted, which is in line with the spectrophotometric measurements. For the maximum thickness of 1.2 mm, the detected intensity reaches approximately 0.75 mW mm$^-2$, which is close to the ChR2 activation threshold of 0.7 mW mm$^-2$ (the red dashed line). Beyond this thickness, the transmitted light intensity may no longer be sufficient to effectively activate the opsins, hence the value of $h_{cyl}=1.2$ mm was chosen for the fabrication of the final device and informed the FEM simulations. Figure \ref{fig:LED_div}d also presents the calculated detected intensity for different LED-to-resin disk distance $d$, using input intensity extracted from the power distribution shown in Figure \ref{fig:LED_div}a. The results show that for LED-to-disk distances $d<5$ mm, the detected intensity exceeds the photostimulation threshold ($I_{det}\geq I_{th}$) across all considered disk thicknesses $h_{cyl}$. This suggests an upper limit of approximately 5 mm for the LED-to-waveguide separation to achieve effective optogenetic stimulation.

With input intensity $I_0=0.13$ mW mm$^-2$, the LED$_{590}$ exhibits a less pronounced attenuation pattern (Figure \ref{fig:LED_div}e). The intensity declines here more gradually, stabilizing at approximately 0.1 mW/mm$^2$ for $h_{cyl}\geq 0.5$ mm. This behavior indicates that at 590 nm, the BMF resin exhibits lower absorption and scattering, leading to a reduced optical attenuation compared to the 470 nm LED.

Our findings demonstrate that the LED$_{470}$ should provide sufficient energy for opsin activation, with a critical constraint on the maximum permissible distance between the light source and the 3D printed waveguide. Specifically, to minimize optical losses and ensure effective opsin activation, this distance should not exceed $\approx 5$ mm (Figure \ref{fig:LED_div}d for $\lambda=470$ nm). These insights are particularly relevant for 3D cell cultures, such as organoids and spheroids, where efficient light penetration is essential to maintain adequate intensity for effective photostimulation.

\subsection{Modeling and 3D-printing of the optogenetic stimulation system}\label{Model}
\begin{figure}
  \includegraphics[width=\linewidth]{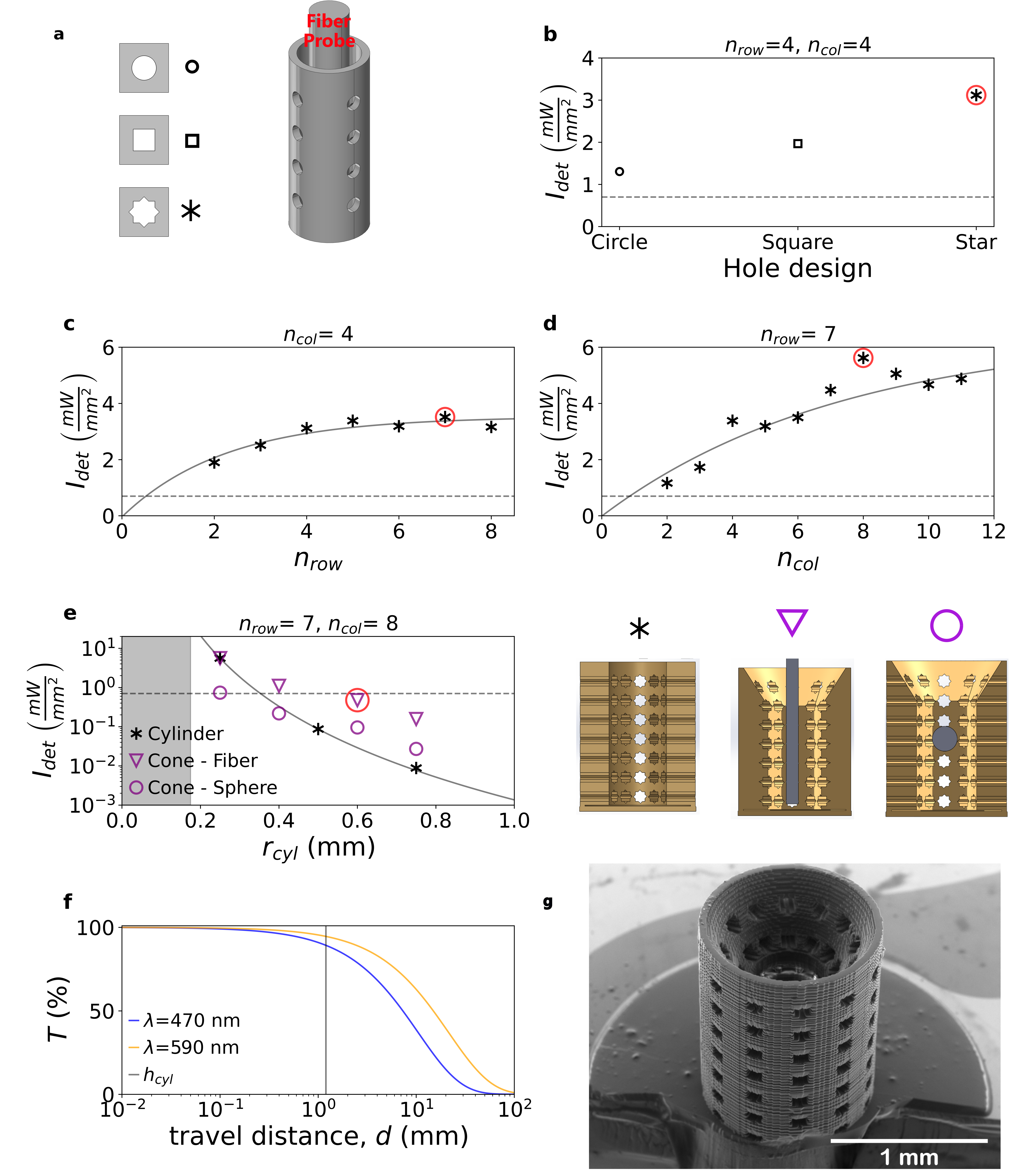}
  \caption{FEM-based device optimization. \textbf{a} Hole geometries considered for the optimization. \textbf{b} $I_{det}$ as a function of three different hole geometries. \textbf{c} $I_{det}$ for star shaped holes as a function of the number of rows $n_{row}$ ($n_{col}=4$ kept constant). \textbf{d} $I_{det}$ as a function of $n_{col}$ ($n_{row}=7$ kept constant). \textbf{e} $I_{det}$ as a function of $r_{cyl}$ ($n_{col}=8$ and $n_{row}=7$ kept constant) for a cylindrical lumen (star markers), and conical lumen with a fiber (purple triangles) and spheroid detector (purple circles). \textbf{f} Transmittance of an infinitely long full cylinder (BMF material) as a function of the ray travel distance $d$ for the two wavelengths of interest. \textbf{g} SEM micrograph of the final fabricated device (see dimensions in the main text).}
  \label{fig:FEM_results}
\end{figure}

The FEM modeling was carried out as shown in \textbf{Figure \ref{fig:FEM_results}}. In the first step, three different hole geometries (circle, square, star, Figure \ref{fig:FEM_results}a) were tested for a fixed number of holes ($n_{row}=n_{col}=4$) and external radius ($r_{cyl}=250$ µm) of the cylinder.
Figure \ref{fig:FEM_results}b shows a comparison of the three different hole geometries for an input intensity $I_0=3.3\cdot10^3$ mW mm $^-2$, indicating that the detected light intensity $I_{det}$ is greater than the threshold intensity $I_{th}\approx0.7$ mW mm$^-2$ that is required for photostimulation of ChR2 \cite{mattis2012principles} (dashed gray horizontal line). The star geometry shows the best performance, followed by the squares and circles. The main rationale is the increased number of discontinuities encountered by the optical rays when interacting with the star shaped openings. Hence, the star geometry was chosen for further optimization. It is worth noting that a vertex number higher than eight for the stars could further improve the light delivery of the 3D printed waveguide, but with a more stringent requirement on the spatial resolution of the printing process. Hence, eight vertices were chosen to simplify the device fabrication process, without compromising the performance in transmittance.

Next, the number of rows, $n_{row}$, was varied along with their vertical spacing (pitch), due to the constraint imposed by the fixed 3D waveguide height $h_{cyl}$ (see Figure \ref{fig:LED_div}d) and the design choice of maintaining equidistant rows. The corresponding FEM results are displayed in Figure \ref{fig:FEM_results}c. $I_{det}$ increases with $n_{row}$ for a small values, followed by a plateau for $n_{row}\geq 5$. This trend arises from two combined effects. First, as more rows are added (with decreasing pitch), the lowest row of holes is positioned closer to the bottom of the 3D printed waveguide, near the incident light source. This reduces absorption losses in the BMF resin before the light reaches the first scattering interface, increasing the amount of power available for redirection toward the center. Second, light that is not scattered at this first stage continues upward, where it encounters additional rows. With more rows and shorter spacing between them, less light is absorbed between the rows, allowing more optical power to reach the detector. Beyond $n_{row}\geq 5$, this effect saturates, indicating that the majority of useful scattering occurs within the first four rows. In Figure \ref{fig:FEM_results}c, the trend is further highlighted by an exponential fitting curve of the form $I_{det}=a+be^{cn_{row}}$, where $a=3.53$, $b=-3.57$, and $c=-0.44$. Based on these findings and with future \textit{in vivo} applications in mind, a value of $n_{row}=7$ was selected (red circle in Figure \ref{fig:FEM_results}c). This choice ensures a sufficient number of outlet pathways for solution containing the targeted cells, especially under configurations where the cell suspension is expected to be introduced from the top of the cylindrical waveguide. In such cases, the side openings facilitate outflow and reduce the risk of clogging, which could otherwise occur due to cell aggregation obstructing the device holes during experiments \cite{Sauret2018}.

In the subsequent step, the effect of the number of columns $n_{col}$ on the total detected intensity was evaluated, starting from the previously optimized value of $n_{row}=7$. As shown in Figure \ref{fig:FEM_results}d, increasing $n_{col}$ beyond 4 led to a further enhancement of $I_{det}$, with gains up to approximately $50\ \%$. This increase is primarily attributed to enhanced scattering of the principal rays by the additional columns in the first row. The effect progressively saturates for $n_{col}\geq 8$. Also in this case, the trend is highlighted by the same exponential fitting curve as for Figure \ref{fig:FEM_results}c, with $a=6.502$, $b=-6.517$, and $c=-0.135$. Based on these results, a design of $n_{col} = 8$ was chosen. 

All the previous simulations were carried out for an external hollow cylinder radius of $r_{cyl}=250$ µm, while keeping the walls thickness constant at $t_{cyl}=50$ µm. Although this compact geometry could support efficient light delivery, it poses practical challenges: the small in-plane dimension can make the alignment of the pyrolytic fiber (diameter of 250 µm) used for the experimental optogenetics characterization extremely difficult (see Section \ref{Opto_test}). To asses how increasing the cylinder radius affects the optical performance, an analysis of the impact of $r_{cyl}$ on the detected light intensity was also carried out. As shown in Figure \ref{fig:FEM_results}e, $I_{det}$ decreases significantly with increasingly larger devices (black star markers); doubling the radius $r_{cyl}$ leads to an intensity drop of nearly two orders of magnitude---falling more than one order below the threshold $I_{th}$). This trend is explained by the angular distribution of scattered rays: as detailed in SI Section~S1, the majority of rays contributing to $I_{det}$ are scattered at zenith angles $\theta_{z}<45$°. When the cylinder radius increases, these rays intersect the central axis at proportionally higher positions, following $z=r_{cyl} \tan(\theta_{z})$. Consequently, a growing number of rays escape the probe's internal volume before reaching the detector, reducing the effective light collection. In contrast, rays scattered perpendicular to the wall ($\theta_{z}=90$°) would intersect the z-axis at a fixed height, making their contribution independent of the in-plane dimensions of the probe.

To compensate for this reduction, the internal wall of the hollow cylinder was designed with a conical geometry, with the top radius being equal to $r_{cyl} - t_{cyl}$, and the bottom to $2/3\ r_{cyl}$ (see schematics in Figure \ref{fig:FEM_results}e, right). The corresponding results are indicated by the purple downward triangles, showing a smaller decrease in detected intensity for increasing $r_{cyl}$ compared to the original design. This is mainly due to the scattering occurring at the first rows of holes, which even in the modified design have a small in-plane distance from the detector. 

Based on this FEM model, the final choice of device to be fabricated consists of: 1) a hollow cylinder with an external radius of $r_{cyl}=0.6$ mm, having a conical internal wall geometry with a bottom radius of 0.4 mm, to make the alignment of the pyrolytic fiber easier (see Section~\ref{Opto_test}); 2) star shaped holes, arranged in 7 rows and 12 columns ($n_{col}$ was increased to compensate for the reduction in $I_{det}$ due to the increase in $r_{cyl}$). The FEM result corresponding to this design is indicated by the red circle in Figure \ref{fig:FEM_results}e with $I_{det}\approx I_{th}$. Working around the stimulation threshold intensity could have the advantage of reducing the excitation volume, better targeting the cells to be stimulated \cite{mattis2012principles}. These results were compared with the case in which an organoid of radius $r_{org}=125$ µm was used as detector (centered within the hollow cylinder at a height of $h_{cyl}/2$ from the bottom, represented by the purple circles in Figure \ref{fig:FEM_results}e, right). $I_{det}$ is one order of magnitude smaller than that observed for the fiber detector. This comparison is meant to stress the importance of accounting for the cellular construct under study. 

For the sake of completeness, light transmission through the BMF printed material as a function of the distance $d$ traveled by the rays inside an infinitely long full cylinder is shown in Figure \ref{fig:FEM_results}f, for which experimental values of $\kappa$ were used as input parameters for the material. It can be observed that, for a travel distance $d=h_{cyl}$ (vertical grey line), the transmission is $T>90 \%$, showing that for the designed device absorption in material should not limit the performances in light delivery.

Figure \ref{fig:FEM_results}g shows an SEM micrograph of the corresponding 3D printed device. As can be observed in the SEM image, the device has significant roughness due to the resolution of the 3D printing process, suggesting that light delivery toward to the center of the hollow cylinder could be further enhanced during actual experiments on photostimulation.

\subsection{Photostimulation of optogenetically modified cells}\label{Opto_test}
\begin{figure}
  \includegraphics[width=\linewidth]{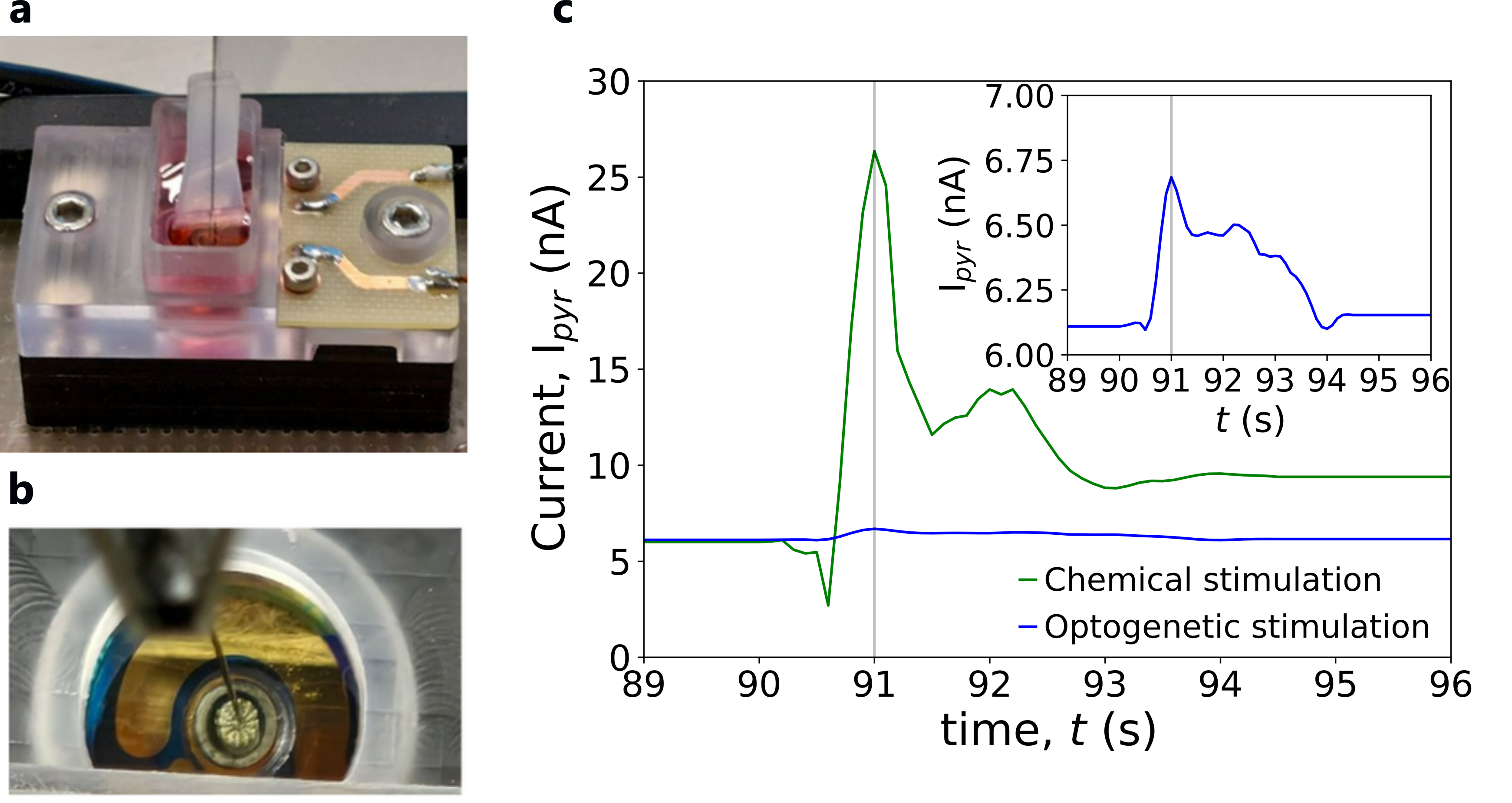}
  \caption{Optogenetics test with the BMF waveguide. \textbf{a} Lateral view of the OpM aligned with the pyrolytic carbon fiber. \textbf{b} Top view of the 3D printed BMF waveguide. \textbf{c} Amperograms obtained at 0.2 mV analyzing of hVM1 cell cultured on top of the pyrolytic fiber. In green--- chemical stimulation via K+; in blue--- optogenetics using LED 470 nm. }
  \label{fig:Pyro_results}
\end{figure}
To evaluate the functionality of the 3D-printed optogenetic system described in Section~\ref{Opt_setup}, we performed photostimulation of optogenetically modified neural stem cells (hVM1-Bcl-X\textsubscript{L}-LChR2-mCherry), combined with amperometric real-time detection of the released dopamine, which based on previous characterization is the main released neurotransmitter \cite{krabbe2009enhanced, courtois2010vitro, vasudevan2019leaky}. The 3D printed waveguide was secured within the custom-fabricated PMMA holder, with a 2 mL liquid reservoir and positioned above the light source  (\textbf{Figure \ref{fig:Pyro_results}a}). A pyrolytic carbon fiber, coated with optogenetically modified cells differentiated for 10 days, was inserted into the central cavity of the optimized 3D-printed waveguide (Figure \ref{fig:Pyro_results}b) \cite{krabbe2009enhanced, courtois2010vitro, vasudevan2019leaky}.
Following approximately  a 90 s of amperometric baseline recording, photostimulation was performed using a 470 nm LED mounted on the optogenetic motherboard, delivering a 500 ms light pulse into the lumen of the 3D printed waveguide. The corresponding current response after optogenetic stimulation (Figure \ref{fig:Pyro_results}c, blue trace), with a magnified view provided in the inset. The time corresponding to the maximal current response is indicated by the grey vertical line. To further assess the viability and functionality of the optogenetically modified cells, dopamine exocytosis was induced via chemical depolarization by elevating the K\textsuperscript{+} concentration of the differentiation medium, in which the 3D printed waveguide and differentiated cells were immersed, to the final concentration of   $\sim 150$ mM. The corresponding current response after chemical stimulation (Figure \ref{fig:Pyro_results}c, green trace) confirms successful real-time detection of dopamine release, validating both the functionality of the optogenetic cells and the efficiency of the acquisition system.
A comparative analysis of the electrical responses elicited by light-induced (blue trace) and chemically induced (green trace) stimulation revealed a marked difference in signal magnitude. While the recorded current amplitude upon chemical stimulation (I\textsubscript{chem}) reached on average about 20 nA, the photostimulation generated a significantly lower current amplitude (I\textsubscript{photo}) of about 600 pA, suggesting suboptimal light delivery by the 3D printed waveguide. The resulting stimulation efficiency (I\textsubscript{photo}/ I\textsubscript{chem}) was only about 2.8\%.   The experimental result is in line with the FEM simulation presented in Figure \ref{fig:FEM_results}f (fiber detector), for which a detected intensity smaller than the threshold (dashed grey horizontal line) was predicted. Other factors may also contribute to this reduced stimulation efficacy. For instance, the culture medium and the 3D microenvironment may introduce optical impedance, further limiting light penetration to the cells. Based on these findings, a further iteration of the 3D waveguide design redesign will be necessary to optimize light delivery and improve the final stimulation efficiency.

\section{Conclusion}
In this study, we have demonstrated the feasibility of fabricating a complex 3D-printed waveguide for optogenetic applications using a projection micro stereolithography (PµSL) printing. The 3D printed waveguide is based on BMF resin, was optimized through geometric refinements and an extensive optical characterization of the resin to enhance light transmission while minimizing losses. In addition, we thoroughly characterized the light source used in this study as its properties significantly influence light propagation and, ultimately, opsin protein activation.

The primary objective of this work was to investigate the optical properties of BMF resin and assess its potential applicability in optogenetics. Our findings provide proof of concept, demonstrating that the 3D-printed BMF waveguide can transmit a sufficient light intensity for optostimulation, resulting in measurable current response during electrochemical real-time detection of the released dopamine. The optogenetic stimulation systems presented in this work is not a fully developed tool for optogenetic applications, but it rather represents an initial investigation into the potential of 3D printed structures for such application. Future work will reduce optical losses by using photobleachable initiator systems and low-absorption fluorinated or cycloaliphatic matrices, further optimizing 3D waveguide geometry, such as vertical alignment offset between successive rows of holes, and minimizing the waveguide–cultured-cell separation, thereby enhancing light transmission and stimulation efficiency 

Ultimately, these advancements will be essential for the future integration of 3D-printed waveguide into complex three-dimensional optogenetic organoid models, supporting further research into disease mechanisms and potential therapeutic strategies.

\medskip
\textbf{Supporting Information} \par %Please delete the Suppporting Information statement if it is not applicable. Please supply Supporting Information in another file. Supporting information should not be provided in .tex format
Supporting Information is available from the Wiley Online Library or from the author.

% Acknowledgements
\medskip
\textbf{Acknowledgements} \par %delete if not applicable))
We thank the HORIZON-EIC-2021-PATHFINDEROPEN program and the projects 4-DBR (Project 101047099) and OpenMIND (Project 101047177), for providing the essential infrastructure and support that made this research possible. The authors acknowledge PolyFabLab, a research infrastructure financially supported by the Novo Nordisk Foundation, grant number NNF21OC0068814.

% References
\medskip

% Use the following code if you wish to generate your bibliography with BibTeX;
% replace the string "MSP-template" below with the name(s) of
% the BibTeX data base(s) you want to use.
% The resulting bibliography-output (the content of the .bbl file)
% must be pasted back into this file before submission.
% Please also include your BibTeX data base file(s) in your submission
% so that we can re-run BibTeX if necessary.
%
\bibliographystyle{MSP}
\bibliography{biblio}

% Please provide Biographies and photos for Essays, Feature Articles, Progress Reports, Reviews, and Perspectives for those authors who should be highlighted  
% These should be at most 100 words long
% For other article types this section can be removed
% Photographs should be 40mm broad and 50 mm high

% \begin{figure}
%   \includegraphics{Figures/KK_picture.png}
%   \caption*{Biography}
% \end{figure}

% \begin{figure}
%   \includegraphics{Figures/KK_picture.png}
%   \caption*{Biography}
% \end{figure}

% \begin{figure}
%   \includegraphics{Figures/KK_picture.png}
%   \caption*{Biography}
% \end{figure}

% \begin{figure}
%   \includegraphics{Figures/KK_picture.png}
%   \caption*{Biography}
% \end{figure}

% Table of contents entry should be 50 - 60 words long
% Image should be 55 mm broad and 50 mm high or 110 mm broad and 20 mm high

\begin{figure}
\textbf{Table of Contents}\\
\medskip
  \centering
  \includegraphics[width=1\linewidth]{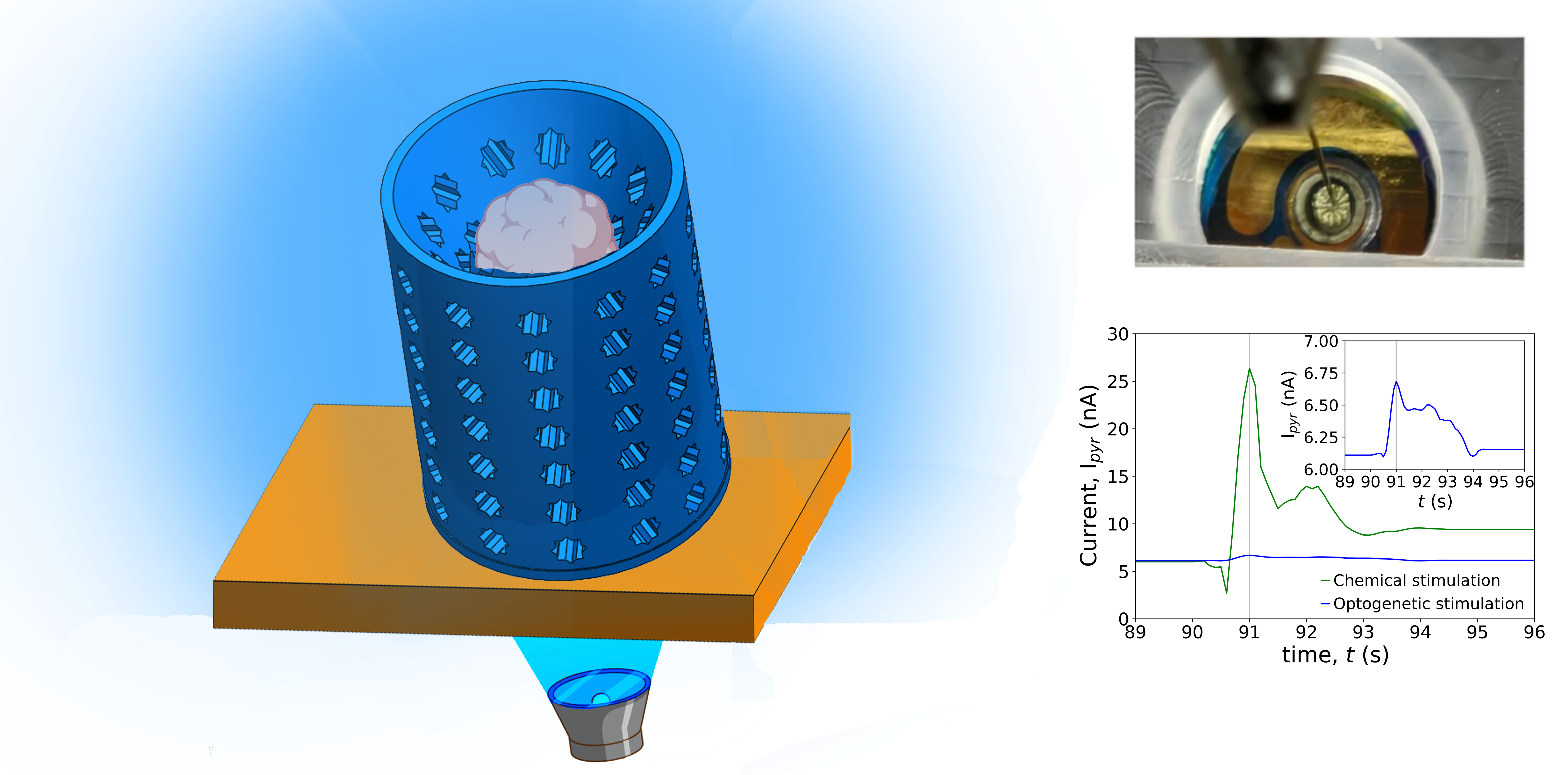}
  \medskip
  \caption*{This article presents the design, optical characterization, and experimental validation of a 3D-printed waveguide for optogenetic stimulation. It details material optical properties, FEM-based design optimization, fabrication of a cylindrical light-delivery system, and functional testing using ChR2-modified dopaminergic cells, demonstrating light-induced dopamine release and outlining future improvements for enhanced stimulation efficiency.}
\end{figure}

\end{document}

\pagestyle{fancy}
\rhead{\includegraphics[width=2.5cm]{vch-logo.png}}

\title{Supporting Information: 3D printed waveguides for optogenetics applications: design optimization and optical characterization}

\maketitle

\author{Giorgio Scordo*}
\author{Kostas Kanellopulos}
\author{Surangrat Thongkorn}
\author{Samuel Tavares da Silva Maraschin}
\author{Kambiz Ghaseminasab}
\author{Evgeniy Shkondin}
\author{Deepshika Arasu}
\author{Stephan Sylvest Keller}
\author{Arto Rainer Heiskanen}
\author{Marta Perez Pereira}
\author{Jenny Emnéus*}

% Affiliations: Please provide adacemic titles (Prof. or Dr.) for all authors where applicable, and include an institutional email address for all corresponding authors
\begin{affiliations}
Dr. G. Scordo\\
Electronic Engineering Department, University of Rome Tor Vergata, Via del Politecnico 1, 00133 Rome, Italy\\
Email Address: giorgio.scordo@uniroma2.it

Dr. K. Kanellopulos\\
TU Wien, Institute of Sensor and Actuator Systems, Gusshausstrasse 27-29, 1040 Vienna, Austria

Dr. S. Thongkorn\\
Department of Biotechnology and Biomedicine, Technical University of Denmark, Kgs. Lyngby, 2800, Denmark

S. T. D. S. Maraschin\\
Department of Biotechnology and Biomedicine, Technical University of Denmark, Kgs. Lyngby, 2800, Denmark

K. Ghaseminasab\\
Department of Biotechnology and Biomedicine, Technical University of Denmark, Kgs. Lyngby, 2800, Denmark

Dr. E. Shkondin\\
Department of Biotechnology and Biomedicine, Technical University of Denmark, Kgs. Lyngby, 2800, Denmark

D. Arasu\\
Hospital Clínic de Barcelona - Fundació de Recerca Cliníc Barcelona - IDIBAPS - University of Barcelona, Barcelona, Catalonia, Spain

Prof. S.S. Keller\\
National Centre for Nano Fabrication and Characterization (DTU Nanolab), Technical University of Denmark, Ørsteds Plads Building 347, Kgs. Lyngby, 2800 Denmark

Dr. A. R. Heiskanen\\
Department of Biotechnology and Biomedicine, Technical University of Denmark, Kgs. Lyngby, 2800, Denmark

Prof. M. P. Pereira\\
The Institute for Molecular Biology of the Autonomous University of Madrid, Madrid, 20049, Spain

Prof. J. Emnéus\\
Department of Biotechnology and Biomedicine, Technical University of Denmark, Kgs. Lyngby, 2800, Denmark\\
Lund University, Department of Chemistry, Center for Analysis and Synthesis, Box 124, 22100, Lund, Sweden\\
Email Address: jenny.emneus@chem.lu.se

\end{affiliations}

\newpage
\tableofcontents

\newpage
\section{FEM Ray Tracing}
The FEM simulations presented in the main text were based on a \textit{Ray Tracing} (RT) study, conducted using the \textit{Geometrical Optics} (GO) module available in COMSOL Multiphysics. The physical parameters used in the simulations---such as geometry and material parameters---are described in the main text. Here, we provide additional computational details and illustrate the typical ray trajectories obtained from the simulations.

In the GO module, the light source was defined through the \textit{Release from Boundary} condition at the bottom surface of the probe (see main text), with a input power $P_0$. Each simulation released $N=2000$ rays per event, with conical emission with angle $\theta=7$° to match the characteristics of the experimental LED$_{470}$. For each wave vector emitted, $N_{w}=10$ rays were launched, resulting in a total of $N\cdot N_{w}=2\cdot 10^4$ rays per simulation. Each ray carried an input power of $P_{r,0}=P_0/(N\cdot N_w)$. 

\begin{figure}[h]
  \includegraphics[width=\linewidth]{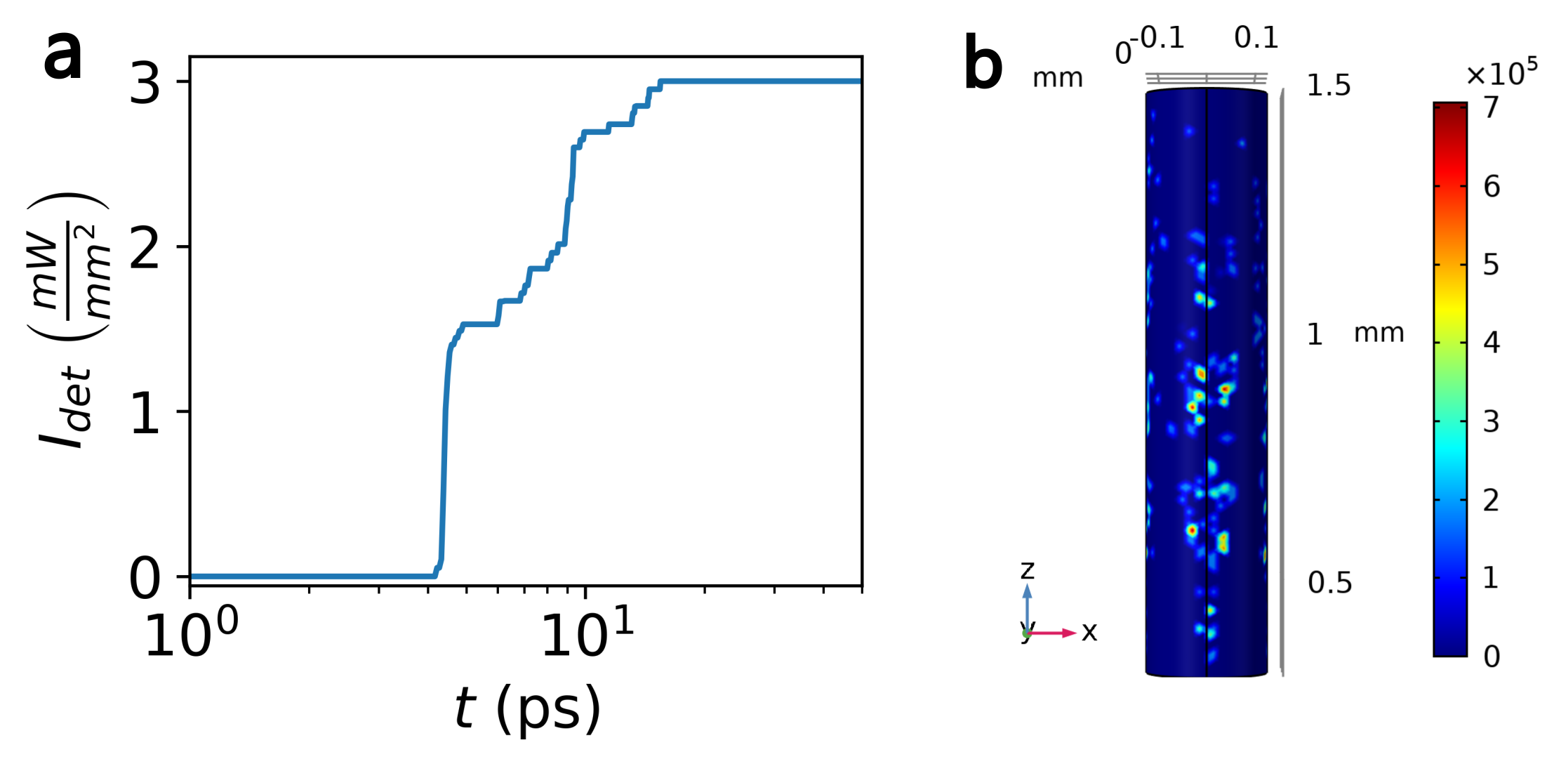}
  \caption{\textbf{a.} Temporal evolution of $I_{det}(t)$.
    \textbf{b.} 3D distribution of the scattered rays impinging onto the detector. The scalebar on the right indicates the corresponding power density $Q_{abs}(\textbf{r}, t=t_0)$, in units [W/m$^2$], at time $t=t_0$.}
  \label{fig:FEM_Idet_t}
\end{figure}

The cylindrical detector was modeled as a \textit{Boundary Wall} set to fully absorb the impinging scattered light rays. For each RT study, light propagation was simulated for a sufficiently long time $t$ to reach steady-state conditions, meaning all rays were either absorbed by the detector or had scattered into free space. An example of this transient behavior is shown in Fig.~\ref{fig:FEM_Idet_t}a, where the total detected intensity $I_{det}$ increases over time, eventually reaching a plateau. The observed staircase profile corresponds to discrete rays arrival at the detector, each contributing to the total signal at different times.
The total detected intensity shown in Fig.~\ref{fig:FEM_Idet_t}a, and discussed in the main text, is obtained upon integration of the absorbed power density $Q_{abs}$ over the detector area $A_{det}$, and normalized by the same quantity,
\begin{equation}
    I_{det}(t) = \frac{1}{A_{det}}\iint_{A_{det}} Q_{abs}(\textbf{r}, t)dA_{det} = \frac{P_{det}(t)}{A_{det}}.
\end{equation}

All values reported in the main text refer to the steady-state value $I_{det}(t\gg0)$.
Fig.~\ref{fig:FEM_Idet_t}b shows the 3D power density distribution $Q_{abs}(\textbf{r}, t=t_0$) across the detector area at a representative time $t_0$. 

\begin{figure}
  \includegraphics[width=\linewidth]{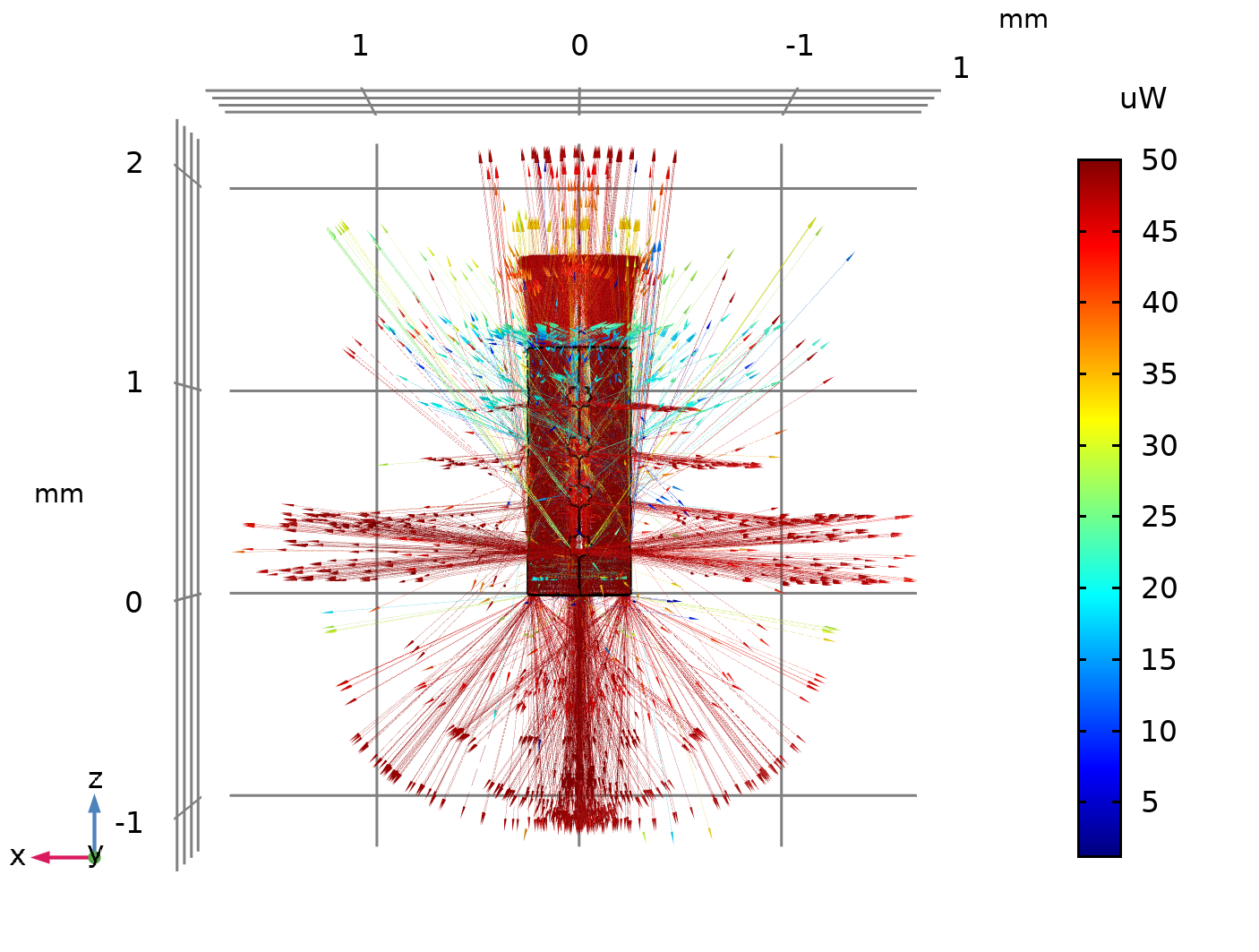}
  \caption{Ray trajectories and corresponding carried optical power $P_{r,sca}$ for $n_{row}=n_{col}=4$, input wavelength $\lambda = 470$ nm, input power $P_0=1$ W, and incident angle $\theta=7$°.}
  \label{fig:FEM_ray_power_yz}
\end{figure}

Fig.~\ref{fig:FEM_ray_power_yz} provides a snapshot of the ray trajectories throughout the simulation domain at the same time $t=t_0$, with the ray color indicating the delivered optical power. In this example, the source power was set to $P_0=1$ W, leading to a maximum power per ray of $P_{r,max}\simeq P_0/(N\cdot N_w)=50$ µW (red rays). The black lines denote the geometry of the probe. 

As shown, the scattered rays form a complex pattern. Rays in the lower region ($z<0$) that propagate downward ($k_z<0$) result primarily from backscattering by the first row of holes, and do not contribute to $I_{det}$. Among the forward-propagating rays ($k_z>0$), some pass directly through the structure (for this specific case, the majority) and are lost into free space, while others interact with the wall of the probe. The outward-propagating rays do not contribute to $I_{det}$ and are similarly lost into free space, whereas only inward-scattered rays---those directed toward the detector---contribute to the final $I_{det}$.

In particular, the outward-propagating rays are scattered with zenith angles $\theta_z\leq90$°, with the first row of holes being the stronger scatterer region. Rays scattered by this first row typically carry power $P_{r, sca}>45$ µW each, while those scattered from higher rows deliver reduced power ($P_{r,sca}\sim 20$ µW) at smaller zenith angles ($\theta_z<90$°). 

\begin{figure}
  \includegraphics[width=\linewidth]{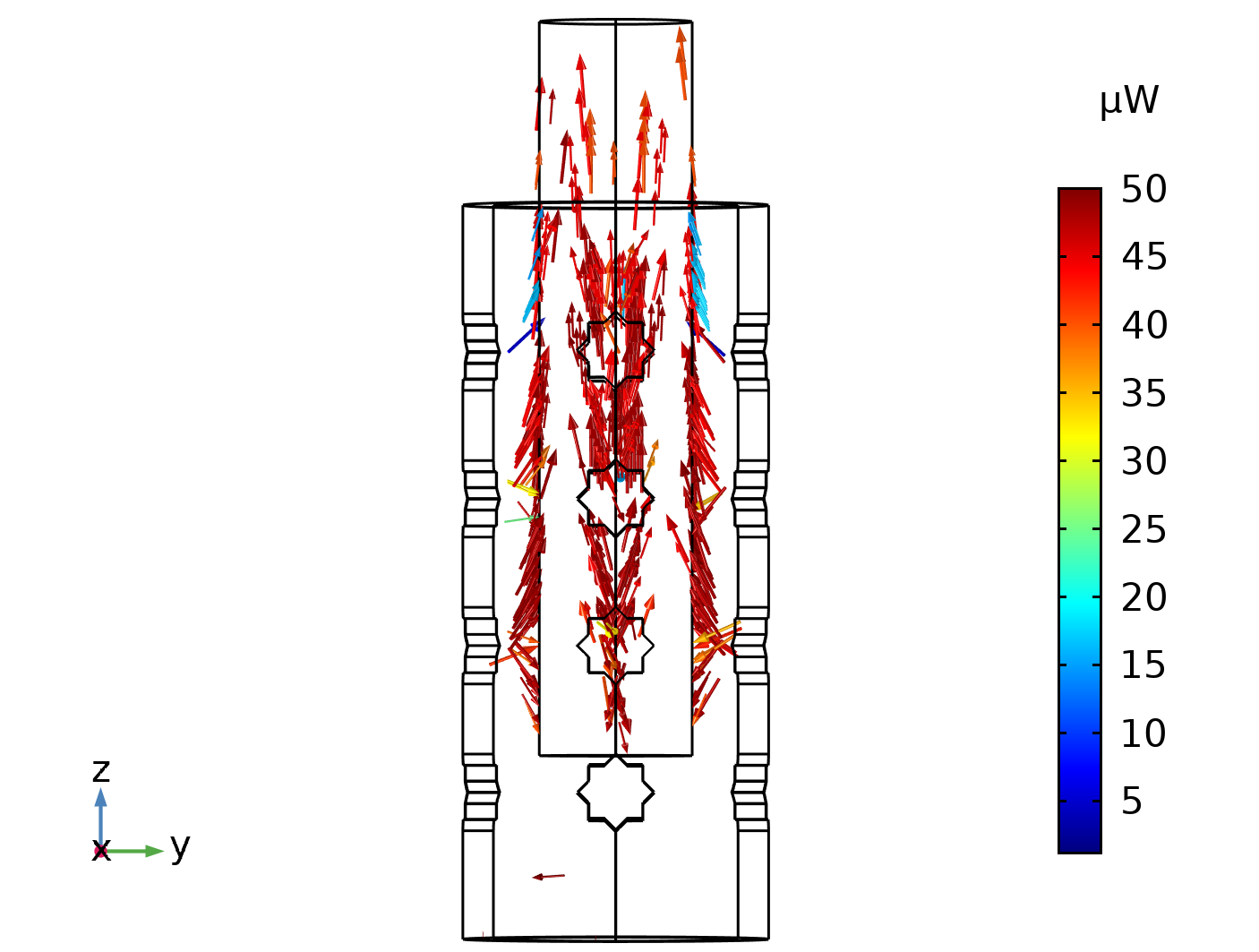}
  \caption{Ray trajectories and corresponding carried optical power $P_{r,sca}$ for $n_{row}=n_{col}=4$, input wavelength $\lambda = 470$ nm, input power $P_0=1$ W, and incident angle $\theta=7$°.}
  \label{fig:FEM_ray_power_xy}
\end{figure}

The inward-scattered rays contributing to $I_{det}$, shown in Fig.~\ref{fig:FEM_ray_power_xy}, are mostly scattered at zenith angles $\theta_z\ll90$°. This directional bias explains the observed reduction in $I_{det}$ as the probe cylinder radius $r_{cyl}$ increases (Fig.~6e in the main text): rays scattered at lower $\theta_z$ intesect the central axis at greater heights ($z=r_{cyl}\tan(\theta_z)$, increasing the likehood of escaping the probe before reaching the detector. This effect is discussed in further details in the main text.

\newpage
\section{BMF optical characterization}
\subsection{Ellipsometry measurements}
Ellipsometry provides a non-destructive means of extracting the complex dielectric permittivity \(\langle\epsilon\rangle\) of a material by analyzing the change in polarization of reflected light. Assuming a single interface (air/material) and a known angle of incidence \(\phi\), the permittivity can be calculated as follows~\cite{fujiwara2007spectroscopic}:
\begin{equation}
    \langle\epsilon\rangle = \sin^2(\phi) \left[1 + \tan^2(\phi) \left(\frac{1 - \rho}{1 + \rho}\right)^2\right],
\end{equation}
where \(\rho\) is the ratio of the Fresnel reflection coefficients for p- and s-polarized light:
\begin{equation}
    \rho \equiv \frac{r_p}{r_s} = \tan(\Psi) e^{i\Delta},
\end{equation}
with \(\Psi\) and \(\Delta\) representing the amplitude ratio and phase difference, respectively, obtained from the ellipsometric measurement.

The complex permittivity is expressed as \(\langle\epsilon\rangle = \epsilon_1 - i\epsilon_2\), from which the refractive index \(n\) and the extinction coefficient \(\kappa\) can be extracted using:
\begin{equation}
    n = \left[\frac{\epsilon_1 + \sqrt{\epsilon_1^2 + \epsilon_2^2}}{2}\right]^{1/2},
\end{equation}
\begin{equation}
    \kappa = \left[\frac{-\epsilon_1 + \sqrt{\epsilon_1^2 + \epsilon_2^2}}{2}\right]^{1/2}.
\end{equation}

\begin{figure}[h]
  \centering
  \includegraphics[width=0.5\linewidth]{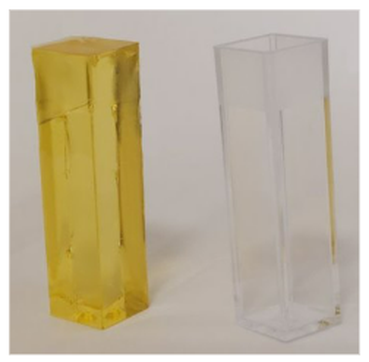}
  \caption{Spectrophotometer samples. Left: 3D printed BMF cuvette 1x1x5 cm. Right: a standard transparent cuvette.}
  \label{fig:cuvette}
\end{figure}

\newpage
\subsection{Data compilation}
The values of refractive index and extinction coefficients used in Fig.~4 in the main text for various commonly used material for optogenetics applications are listed in Table~\ref{tab:mat}.

\begin{table}[h]
\centering
 \caption{List of the refractive index $n$ and extinction coefficient $\kappa$ for $\lambda = 470, 590$ nm.}
 \label{tab:mat}
  \begin{tabular}[htbp]{@{}lllll@{}}
    \hline
    Material & $\lambda$ (nm) & $n$ & $\kappa$ & Ref. \\
    \hline
    SiN & 470 & 2.05 & $2.5\cdot 10^{-6}$ & \cite{corato2024absorption} \\
    & 590 & 2.023 & $3.57\cdot 10^{-7}$ & \cite{corato2024absorption} \\
    \hline
    Si & 470 & 4.485 & $7.02\cdot10^{-2}$ & \cite{Schinke2015} \\
     & 590 & 3.954 & $2\cdot10^{-2}$ & \cite{Schinke2015} \\
     \hline
    SiO$_2$ & 470 & 1.477 & $1.5\cdot10^{-5}$ & \cite{Franta2016} \\
    & 590 & 1.471 & $3.73\cdot10^{-6}$ & \cite{Franta2016} \\
    \hline
    SiC & 470 & 3.47 & 0.51 & \cite{Larruquert2011} \\
    & 590 & 3.41 & 0.32 & \cite{Larruquert2011} \\
    \hline
    SU8 & 470 & 1.6 & $3.74\cdot10^{-6}$ & \cite{chen2024implantation}\\
    & 590 & 1.59 & $3.03\cdot10^{-6}$ & \cite{chen2024implantation}\\
    \hline
    PAG & 470 & 1.5 & $1.72\cdot10^{-6}$ & \cite{monney2021photolithographic} \\
     & 590 & 1.5 & $1.08\cdot10^{-6}$ & \cite{monney2021photolithographic} \\ 
    \hline
    PEGDA & 470 & 1.45 & $4.49\cdot10^{-8}$ & \cite{feng2020printed}\\
    & 590 & 1.45 & $5.63\cdot10^{-9}$ & \cite{feng2020printed} \\
    \hline
    PDMS & 470 & 1.41 & $1.43\cdot10^{-6}$ & \cite{Zhang2020} \\
    & 590 & 1.40 & $1.54\cdot10{-6}$ & \cite{Zhang2020} \\
    \hline
    FormLabs Clear & 470 & 1.518 & $0.2\cdot 10^{-6}$ & $n$ \cite{reynoso2021refractive}, $\kappa$ \cite{bartels2024geometrical}\\
     & 590 & 1.509 & $0.14\cdot 10^{-6}$ & $n$ \cite{reynoso2021refractive}, $\kappa$ \cite{bartels2024geometrical}\\
    \hline
    OrmoComp & 470 & 1.53 & $0.19\cdot 10^{-6}$ & $n$ \cite{gissibl2017refractive}, $\kappa$ \cite{jost2024fabrication}\\
     & 590 & 1.52 & $9.4\cdot10^{-8}$ & $n$ \cite{gissibl2017refractive}, $\kappa$ \cite{jost2024fabrication}\\
    \hline
  \end{tabular}
\end{table}

\newpage
\section{Optical motherboard}
\subsection{Schematic circuit}
\begin{figure}[h]
  \centering
  \includegraphics[width=0.4\linewidth]{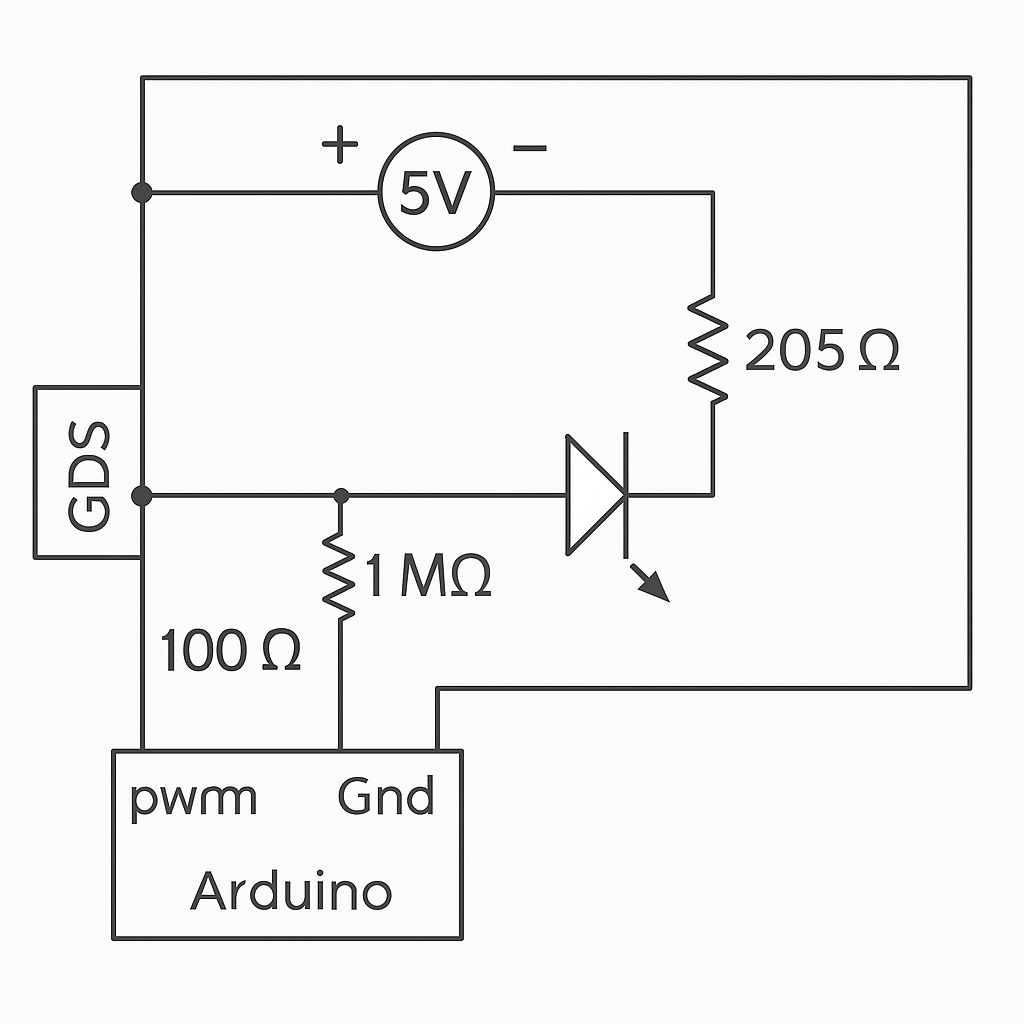}
  \caption{The schematic diagram of the equivalent circuit used for driving the LED}
  \label{fig:LED_circuit}
\end{figure}

\begin{figure}[h]
  \centering
  \includegraphics[width=0.6\linewidth]{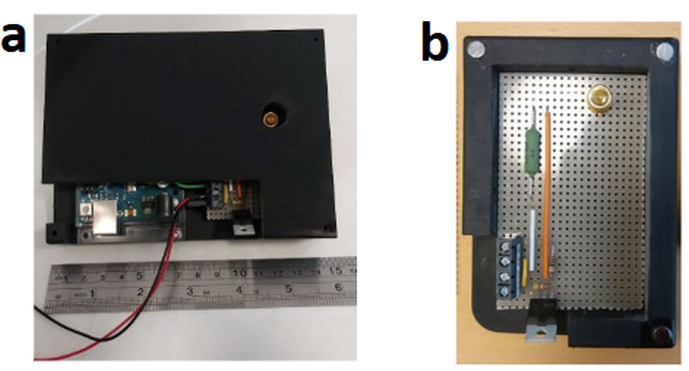}
  \caption{\textbf{a} Full assembled opto-motherboard device (OpM). \textbf{b} Top view of the OpM LED side.}
  \label{fig:OpM}
\end{figure}

The designed Optical Motherboard comprises electrical transmission lines embedded in a matrix board (RS Components, UK), as shown in Fig.~\ref{fig:OpM}b, where LED$_{470}$ or LED$_{590}$ are mounted---a schematics of the LED circuit is shown in Fig.~\ref{fig:LED_circuit}. A microcontroller board, Arduino Uno (Italy) regulates the current and controls pulse stimulation via a dedicated script (section \ref{Sec:script}). The OpM is enclosed within a 3D-printed shell fabricated from black tough resin using a Form 3 printer (Formlabs, USA). Neodymium magnets from RS Components, UK, are embedded along the edges to facilitate the attachment of interchangeable lids, enabling a modular configuration.

\newpage
\subsection{Arduino script}\label{Sec:script}
The script used to activate the LED implements a loop in which the LED is turned on for six cycles of 500 ms, followed by an off period of 50 s.

\begin{lstlisting}[caption={Arduino sketch for 4-DBR blinking LED}, label={lst:arduino_led}]
/*
  Blinking LED

  model, check the Technical Specs of your board at:
  https://www.arduino.cc/en/Main/Products
  https://www.arduino.cc/en/Tutorial/BuiltInExamples/Blink
*/

// the setup function runs once when you press reset or power the board
void setup() {
  // initialize digital pin 9 as an output.
  pinMode(9, OUTPUT);
  digitalWrite(9, LOW);
  delay(60000);
}

// the loop function runs over and over again forever
void loop() {
  digitalWrite(9, HIGH);  // turn the LED on (HIGH is the voltage level)
  delay(500);
  digitalWrite(9, LOW);   // turn the LED off
  delay(50000);

  digitalWrite(9, HIGH);  // on
  delay(500);
  digitalWrite(9, LOW);   // off
  delay(50000);

  digitalWrite(9, HIGH);
  delay(500);
  digitalWrite(9, LOW);
  delay(50000);

  digitalWrite(9, HIGH);
  delay(500);
  digitalWrite(9, LOW);
  delay(50000);
}
\end{lstlisting}

\subsection{Light cavities on Si wafer and 3D printing parameters}
\begin{figure}[h]
  \centering
  \includegraphics[width=1\linewidth]{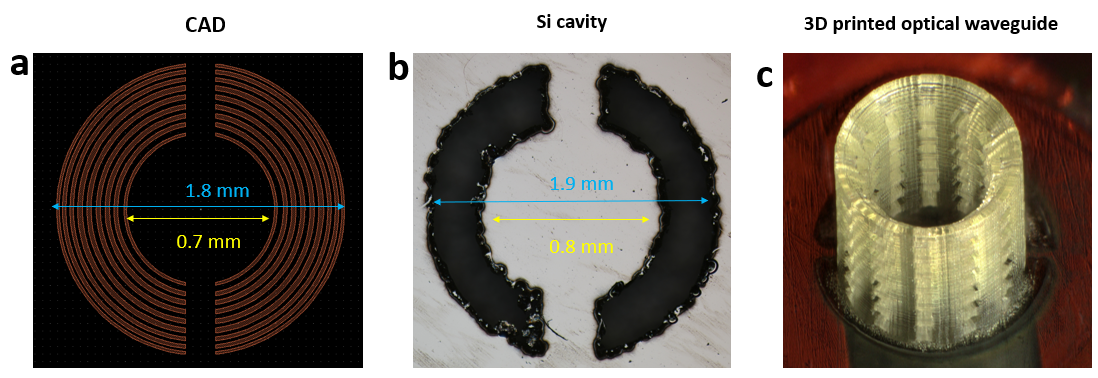}
  \caption{Fabrication process of the optical cavity: \textbf{a.} CAD design consisting of multiple semicircular lines used as the laser path. \textbf{b.} Top view of the fabricated optical cavity obtained after 5000 repetitions. \textbf{c.} Isometric view of the 3D-printed optical waveguide aligned on top of the cavity.}
  \label{fig:cavity}
\end{figure}

To obtain two identical openings for positioning the 3D-printed probe on top of the light source, two symmetrical semicircular apertures were first designed using CAD software Layout Editor (Fig.~\ref{fig:cavity}a). The silicon chips containing gold electrodes were then processed using a laser micromachining system (3D-Micromac AG, Germany) operated at a power density of 7 mW/mm, a scanning speed of 250 µm/s, and 5000 repetitions. This procedure produced precise apertures on both sides of the electrode (Fig.~\ref{fig:cavity}b). Subsequently, the chips were cleaned with ethanol to remove residual dust and debris.

To fabricate the 3D optical waveguide on the Si cavity, the ChituBox software (Chitu Systems, China) was employed for print preparation, using a layer thickness of 5 µm and fabricated using a Boston Micro Fabrication (BMF) projection microstereolithography (PµSL) 3D printer.
The printing procedure was divided into three sequential exposure stages to ensure proper adhesion, structural stability, and complete polymerization of the printed microstructure. 
In the first stage, a single base layer was exposed for 1 s at 40$\%$ light intensity (corresponding to an effective optical power of  37.6 mW/cm2). This layer served to promote strong adhesion of the photocurable resin to the silicon chip surface and establish a uniform foundation for subsequent layers.
In the second stage, the following 49 layers were cured under a 1 s exposure at 30$\%$  light intensity ( 28.2 mW/cm2 ) with a 2 s interlayer delay. This step allowed gradual structure buildup while minimizing internal mechanical stress and avoiding over-curing, which can cause deformation in high-aspect-ratio geometries.
In the final stage, the remaining 261 layers were polymerized under a 1 s exposure at 45$\%$ ( 42.3 mW/cm2 )intensity with a 60 s interlayer delay.
The resulting microstructure exhibited uniform layer stacking and smooth interfaces, confirming the precision and stability of the PµSL printing process under these optimized exposure conditions (Fig.~\ref{fig:cavity}c). 

\section{2D electromechanical sensor fabrication}
Below is the detailed cleanroom step-by-step protocol used for device preparation.

\subsubsection*{1. Base Wafer Preparation}

\begin{itemize}
  \item \textbf{1.1 Stock-out} – 150\,mm Si \textless100\textgreater, 675\,\textmu m thick, single-side polished, n-type (P).
\end{itemize}

\subsubsection*{2. Nanograss Template on Si}

\begin{itemize}
  \item \textbf{2.1 ICP Etch (Pegasus 2)} – Run “CORE” recipe to generate nanograss texture (\cite{Nguyen_2020}).
  \item \textbf{2.2 RCA Clean} – Sequential RCA-I and RCA-II.
  \item \textbf{2.3 Wet Oxidation} – Anneal+Oxide furnace (C1), WET at 1100\,\textdegree C for 80\,min (full boat), targeting 600\,nm SiO\textsubscript{2} thickness.
\end{itemize}

\subsubsection*{3. Spin-Coating SU-8 2035 (\textasciitilde15\,\textmu m)}

\begin{itemize}
  \item \textbf{3.1 Syringe Preparation} – In fume hood, fill syringes with SU-8 2035$\geq$24 h before use; record bottle lot and expiry.
  \item \textbf{3.2 Plasma Descum} – Plasma Asher 2: 400\,sccm O\textsubscript{2}, 70\,sccm N\textsubscript{2}, 1\,kW, 75\,min.
  \item \textbf{3.3 Spin-Coating} – RC08, program “BioMic 2035 15\,\textmu m (6-inch)”:
    \begin{itemize}
      \item 1000\,rpm for 10\,s (200\,rpm/s) $\rightarrow$ 5000\,rpm for 120\,s (1000\,rpm/s), \textit{manual dispense}.
    \end{itemize}
  \item \textbf{3.4 Soft-Bake} – Hot plate 1273: 50\,\textdegree C for 15\,min, with 2\,\textdegree C/min ramps. Cool for 30\,min.
  \item \textbf{3.5 Backside Wipe} – Clean SU-8 from wafer backside using acetone and cleanroom tissue. Also wipe the hot plate.
\end{itemize}

\subsubsection*{4. SU-8 Exposure and PEB}

\begin{itemize}
  \item \textbf{4.1 UV Exposure} – MA6-2, soft contact, 2\,$\times$\,250\,mJ/cm\textsuperscript{2}.
  \item \textbf{4.2 Post-Exposure Bake (PEB)} – Hot plate 1273: 50\,\textdegree C for 2\,h with 2\,\textdegree C/min ramps.
\end{itemize}

\subsubsection*{5. SU-8 Development, Hard-Bake, and Characterization}

\begin{itemize}
  \item \textbf{5.1 Development} – PGMEA (NICOSTO), three 1-min puddles on Gamma spin-coater.
  \item \textbf{5.2 Flood Dose Test} – MA6-2, 2\,$\times$\,250\,mJ/cm\textsuperscript{2}, 30\,s wait.
  \item \textbf{5.3 Hard-Bake} – 90\,\textdegree C for 15\,h with 2\,\textdegree C/min ramps.
  \item \textbf{5.4 Optical Inspection} – Microscope inspection of dose-test features.
  \item \textbf{5.5 Thickness Measurement} – Dektak/XTA; target: 15\,\textmu m. If deviation >15\%, check resist age; note edge bead thickening.
\end{itemize}

\subsubsection*{6. Pyrolysis of SU-8 to Glassy Carbon}

\begin{itemize}
  \item \textbf{6.1 Furnace Preparation} – Run \texttt{BioMic\_Prep\_N2} unless furnace was used $<$72\,h ago.
    \begin{itemize}
      \item Load: wafers face-up, major flat up, boat fully to rear; front side faces N\textsubscript{2} flow.
    \end{itemize}
  \item \textbf{6.2 Pyrolysis} – Use recipe \texttt{nicostot3-step\_LowRamp\_1050C\_2h}.
  \item \textbf{6.3 Post-Run Clean} – Run \texttt{BioMic\_CleanAfterUsed\_ActiveCooling}.
  \item \textbf{6.4 Carbon Thickness} – Dektak/XTA; aim for 2.2–2.3\,\textmu m (same >15\% deviation rule as in 5.5).
  \item \textbf{6.5 Optical Check} – Microscope inspection of dose-test patterns.
  \item \textbf{6.6 Sheet Resistance} – 4-point probe on test structures.
  \item \textbf{6.7 SEM Cleanliness Check} – Supra 2 (InLens/SE2); verify no dust or metal.
\end{itemize}

\subsubsection*{7. Metal-Contact Photolithography (AZ 5214E, 4.2\,\textmu m)}

\begin{itemize}
  \item \textbf{7.1 Spin-Coating} – Gamma coater, program \texttt{6141-DCH}, HMDS prime, 3\,$\times$\,11.4\,s coating, 20\,s wait $\rightarrow$ 4.2\,\textmu m film.
  \item \textbf{7.2 Pattern Exposure} – MA6-2, contact mode, exposure time 20\,s (design-dependent).
  \item \textbf{7.3 Reversal Bake} – 110\,\textdegree C for 180\,s.
  \item \textbf{7.4 Flood Exposure} – MA6-2, 20\,$\times$\,49.5\,mJ/cm\textsuperscript{2} ($\approx$1\,J/cm\textsuperscript{2}), 4.5\,s exposure, 20\,s wait.
  \item \textbf{7.5 Development} – TMAH (1007 DCH), 5\,$\times$\,60\,s steps; extend in 60\,s increments until clear and uniform.
  \item \textbf{7.6 Final Inspection} – Dektak + microscope; verify exposure quality, alignment, resist thickness, and coverage.
\end{itemize}

\bibliographystyle{MSP}
\bibliography{biblio}